# Structure and Dynamics of the Young Massive Star Cluster Westerlund 1

Lingfeng Wei (魏凌枫)[1], Peter C. Boyle[2], Jessica R. Lu[3], Matthew W. Hosek, Jr.[4], Quinn M. Konopacky[5], Richard G. Spencer[6], Dongwon Kim[7], Nicholas Z. Rui[8], Max Service[9], D. B. Huang[10], and Jay Anderson[11]

[1] Department of Physics, University of California, San Diego, La Jolla, CA 92093, USA; wei-lingfeng@ucsd.edu
[2] Department of Radiation Oncology, University of California, Los Angeles, Los Angeles, CA 90095, USA
[3] Department of Astronomy, University of California, Berkeley, Berkeley, CA 94720, USA
[4] Department of Physics and Astronomy, University of California, Los Angeles, Los Angeles, CA 90095, USA
[5] Department of Astronomy and Astrophysics, University of California, San Diego, La Jolla, CA 92093, USA
[6] Affiliate, Department of Mathematics, University of Maryland, College Park, MD 20742, USA
[7] Cinamon Corp., 484 Gangnam-daero, Gangnam-gu, Seoul 06120, Republic of Korea
[8] TAPIR, California Institute of Technology, Pasadena, CA 91125, USA
[9] W. M. Keck Observatory, 65-1120 Mamalahoa Highway, Kamuela, HI 96743, USA
[10] Verkada Inc., 406 East 3rd Avenue, San Mateo, CA 94401, USA
[11] Space Telescope Science Institute, Baltimore, MD 21218, USA

Received 2025 January 28; revised 2025 August 11; accepted 2025 August 11; published 2025 October 17

## Abstract

We present a structural and dynamical analysis of the young massive star cluster Westerlund 1 (Wd1). Using multiepoch Hubble Space Telescope observations, we measure the proper motions of 10,346 stars and present an unprecedented determination of cluster membership probability. We then determine the color membership after correcting for extinction. The stellar density map weighted by the membership probability and completeness correction displays a spatial elongation aligned with the Galactic plane with a high eccentricity of 0.71. The radial stellar density profile shows a decreasing core radius with increasing mass, indicative of weak but detectable mass segregation. Quantitatively, the measured mass segregation ratio is $\Lambda_{MSR} = 1.11 \pm 0.11$, only above unity by $1\sigma$. Building on the structural modeling along with a suite of stellar evolutionary and atmospheric models, we fit the cluster age and distance in two color–magnitude diagrams, concluding that Wd1 is approximately $7.45 \pm 0.53$ Myr old and at a distance of $3.7 \pm 0.1$ kpc, with a degeneracy in the posterior which is compatible with the literature values. We measure a 1D velocity dispersion of $4.13 \pm 0.13$ km s$^{-1}$, indicating a subvirialized state. The crossing time is 0.30 Myr and the relaxation time is 0.26 Gyr. Given the age of Wd1, we expect dynamical mass segregation for stars more massive than 12 $M_\odot$, which accounts for the minor mass segregation observed in the mass range of 1.12–15.21 $M_\odot$ in this work. This suggests that the mass segregation in Wd1 is more likely a dynamical effect.

*Unified Astronomy Thesaurus concepts:* Star clusters (1567); Star formation (1569); Star forming regions (1565); Stellar kinematics (1608)

*Materials only available in the online version of record:* machine-readable table

## 1. Introduction

Young massive clusters (YMCs) are areas of intense star-forming activity and offer unique insight into star formation, cluster modeling, and the initial mass function (IMF; S. F. Portegies Zwart et al. 2010). YMCs exist within disparate environments, ranging from the Galactic center to the Galactic disk. Investigating differences in their physical properties allows us to determine the dependence of star formation on initial conditions. Finding similarities would indicate which fundamental mechanisms prevail despite such perturbations.

Westerlund 1 (Wd1) is one of the most massive young star clusters in the Galaxy. Discovered more than half a century ago (B. Westerlund 1961), Wd1 is an ideal site to study a starburst environment in detail for its youth, proximity, and rich population of stars across a wide range of masses. Substantial efforts have been devoted to accurately determining the fundamental properties of Wd1, such as its age and distance, to improve our understanding of this cluster.

Recent studies based on Gaia EDR3 and eclipsing binary analysis have converged on a distance estimate for Wd1 of 4.0–4.3 kpc (E. R. Beasor et al. 2021; F. Navarete et al. 2022; I. Negueruela et al. 2022). In contrast, M. Aghakhanloo et al. (2020) and M. Aghakhanloo et al. (2021) reported a significantly closer distance of 2.6–2.8 kpc using Gaia Data Release 2 parallax measurements, although these estimates are potentially subject to sample selection bias (I. Negueruela et al. 2022). With Hubble Space Telescope (HST) optical and infrared (IR) observations, M. W. Hosek et al. (2018) reported a distance ranging from 4.1 to 5.2 kpc depending on the cluster age.

The diverse stellar populations in Wd1 enabled a variety of proxies for cluster age estimation. Previously, J. S. Clark et al. (2005) and P. A. Crowther et al. (2006) independently concluded a cluster age of 3.5–5 Myr according to the formation models of Wolf–Rayet (W-R) stars and supergiants. I. Negueruela et al. (2010) reported an age of 6.3 Myr with isochrones fitting B supergiants at 4.5–5 kpc. E. R. Beasor et al. (2021) determined a pre-main-sequence (PMS) age of 7.2 Myr, while the eclipsing binary W13 sets a minimum age of 5 Myr, and the faintest red supergiants (RSGs) suggest an older age of 10.4 Myr. These age estimates were further supported by F. Navarete et al. (2022) and D. F. Rocha et al. (2022), in which the authors proposed that the broad age spread may reflect multiple episodes of star formation within the cluster.

Accurate measurements of the cluster age and distance from color–magnitude diagrams (CMDs) require modeling the kinematic and photometric membership and an extinction map.

As one of the most massive young clusters in the Milky Way, Wd1 is a benchmark for testing models of star cluster formation and evolution. Due to the limited photometric depth of previous studies, the initial cluster mass of Wd1 has been mostly derived by extrapolating a canonical IMF (J. S. Clark et al. 2005; W. Brandner et al. 2008; M. Andersen et al. 2017). This method suggested a cluster mass of $4$–$6 \times 10^4\ M_\odot$, corresponding to a virial equilibrium velocity dispersion of 4–6 km s$^{-1}$. This estimate significantly exceeds the previously measured velocity dispersion of $2.1^{+3.4}_{-0.9}$ km s$^{-1}$ (M. Cottaar et al. 2012), implying that Wd1 is subvirial and gravitationally bound, despite having dispersed most of its gas mass at an age of $\sim 7$ Myr. The virial state directly impacts these models, influencing assumptions about initial conditions, gas expulsion, and feedback in massive clusters.

Uncertainties about the virial state of Wd1 persist due to the reliance on assumptions about its IMF. Currently, there are conflicting results on the high-mass slope of the IMF of Wd1, e.g., $\alpha = -2.44^{+0.08}_{-0.20}$ in the mass range 3.5–27 $M_\odot$ by M. Gennaro et al. (2011) and $\alpha = -1.8 \pm 0.1$ in the mass range 5–100 $M_\odot$ by B. Lim et al. (2013). For low-mass stellar content, M. Andersen et al. (2017) argued for the IMF in the cluster to be consistent with the canonical IMF for the mass range 0.15–1.4 $M_\odot$. Recent studies have reported similarly unusual IMFs in other YMCs, such as the Galactic center (J. R. Lu et al. 2013) and the Arches cluster (M. W. Hosek et al. 2019), raising the question of whether these deviations are environmentally driven or are intrinsic to YMCs in general. Wd1 provides an excellent opportunity to test this. Kinematic membership is critical to remove field star contamination, which could be high enough to inflate the slope of the IMF in the low-mass and substellar regimes.

Mass segregation in Wd1 offers critical insights into the cluster's dynamical evolution and star formation history. However, contradictory results have also been reported on the mass segregation in Wd1. M. Gennaro et al. (2011) argued that the cluster appears mass segregated, based on the radial variations of the IMF slope. M. Gennaro et al. (2017) later revisited the cluster using the mass segregation ratio $\Lambda_{MSR}$, and found little evidence of mass segregation for stars more massive than 3.5 $M_\odot$, except for the most massive stars above 40 $M_\odot$. The authors concluded that Wd1 was not primordially segregated. The different result from M. Gennaro et al. (2011) is attributed to fitting IMF slopes involving rather arbitrary binning, which could bias toward a high degree of mass segregation (M. Cottaar et al. 2012; M. Gennaro et al. 2017).

In this paper, we determine the kinematic and photometric cluster members of Wd1 using observations with multiple epochs and filters from the HST. With an extinction map based on the main-sequence (MS) population in the cluster, we present its differentially dereddened, field-decontaminated CMDs in the mass range 1.12–15.21 $M_\odot$. We derive the radial profile of Wd1 and examine the degree of mass segregation using the cluster members.

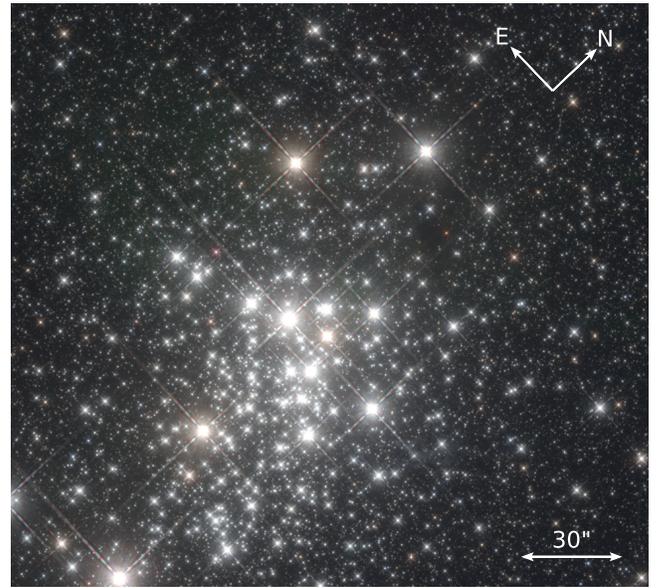

**Figure 1.** An HST WFC3-IR image of Wd1 in false color. A logarithmic color stretch is applied, with the F160W, F139M, and F125W images mapped to red, green, and blue, respectively, after independent normalization of each channel.

This paper is structured as follows. Section 2 illustrates the observation details. Section 3 introduces the extraction and calibration of the data. Section 4 elaborates on the methods we use for modeling the properties of the cluster, including determining cluster membership for each source, the extinction map, and completeness. In Section 5, we present the main results, including the surface density, morphology, radial density profile, cluster age and distance, radial profile, velocity dispersion, and mass segregation. We discuss the elongation, virial state, radial profile comparisons, dynamical timescales, the origin of mass segregation, and their implications in Section 6. Finally, we summarize the conclusions in Section 7.

## 2. Observations

Wd1 (R.A. = 16$^h$47$^m$05$^s$57, decl.=$-45°$ 50$'$ 24$''$.14, J2000) was observed with the HST at multiple epochs to measure proper motions (PMs) and in multiple filters to obtain multicolor photometry. The cluster was observed at four different epochs—2005, 2010, 2013, and 2015—and in four filters at both red optical and IR wavelengths. See Figure 1 for a composite image.

The earliest data set was obtained on 2005 January 23 with the Advanced Camera for Surveys (ACS) Wide Field Camera (WFC; GO-10172; R. de Grijs 2004) using the F814W filter ($\lambda = 806$ nm, $\Delta\lambda = 287$ nm). The total exposure time was 2407 s, comprised of three images with small dithers to cover the chip gap. The final image covers a 211$''$ × 218$''$ field of view (FOV). In these individual optical exposures, stars brighter than $m_{F814W} = 18.4$ were saturated; however, astrometry and photometry can still be extracted for stars as bright as $m_{F814W} = 13$ with increased uncertainty.

A second data set was obtained in 2010 with the IR channel of the Wide Field Camera 3 (WFC3) using three filters (GO-11708; M. Andersen 2009). Due to the limited FOV of WFC3-IR, $\sim 130''$, a 2 × 2 mosaic was used to cover the entire 2005 ACS-WFC FOV. The three IR filters were F125W ($\lambda = 1249$ nm, $\Delta\lambda = 285$ nm), F139M ($\lambda = 1384$ nm, $\Delta\lambda = 64$ nm), and F160W ($\lambda = 1537$ nm, $\Delta\lambda = 268$ nm). Seven

**Table 1**
Westerlund 1 Observations from the Hubble Space Telescope

| Date[a] | Filter | P.A. (deg) | $t_{\rm exp}$[b] (s) | $N_{\rm img}$[c] | $N_{\rm dith}$ | $N_{\rm stars}$ | $\sigma_{\rm trans}$ (mas) | HST ID |
|---|---|---|---|---|---|---|---|---|
| 2005.485 | F814W | 46.43 | 802 | 3 | 1 | 10,056 | 0.30 | 11708 |
| 2010.652 | F125W | −45.87 | 349 | 7 | 2 × 2 | 10,029 | 0.91 | 11708 |
| 2010.652 | F139M | −45.87 | 899 | 7 | 2 × 2 | 10,028 | 0.92 | 11708 |
| 2010.652 | F160W | −45.87 | 299 | 7 | 2 × 2 | 10,056 | 0.85 | 13044 |
| 2013.199 | F160W | 134.67 | 299 | 14 | 2 × 2 | 10,056 | 0.89 | 13044 |
| 2013.202[d] | F160W | 134.67 | 8 | 1 | 8 × 8 | 6571 | 0.19 | 13044 |
| 2015.148 | F160W | 134.67 | 249 | 13 | 2 × 2 | 10,056 | 0.86 | 13809 |
| 2015.149[d] | F160W | 134.67 | 8 | 1 | 8 × 8 | 6571 | 0.18 | 13809 |

**Notes.**
[a] Images were taken over the course of 2 days. The date used in the PM analysis is the average over the individual images.
[b] Exposure time for a single image.
[c] Number of images at each dither position.
[d] These data were subarrayed to one-quarter of the detector.

images per filter were observed at each point in the mosaic. The total exposure times for each tile were 2444 s in the F125W filter, 6294 s in the F139M filter, and 2094 s in the F160W filter, where each tile comprised seven images.

A third and fourth data sets were obtained in 2013 (GO-13044; J. R. Lu et al. 2016a) and 2015 (GO-13809; J. R. Lu et al. 2016b) with WFC3-IR in the F160W filter to provide multiepoch astrometry for sources detected at IR wavelengths. The position angle of the 2013 and 2015 WFC-IR data was rotated ∼180° with respect to the position angle of the 2010 WFC3-IR data. Two short 8 s exposures were added to provide further information on the brightest stars in case the saturation and errors were unacceptable in the long exposures. Details of the exposure times, number of exposures, and sensitivity of each filter are presented in Table 1.

The data were reduced using the standard online HST data reduction pipeline, and the resulting FLT images (`*_flt.fits`) were downloaded from the HST archive. All astrometric and photometric measurements were extracted from the individual FLT images and then combined as described in Section 3.1. The drizzle-combined DRZ images (`*_drz.fits`) for each filter and epoch were used to calibrate the photometric zero-points.

## 3. Data Analysis

### 3.1. Astrometric and Photometric Extraction

We construct an astrometric and photometric catalog for each HST data set as follows:

i. extract star lists of stellar positions and fluxes for each image;
ii. crossmatch the star lists and transform them into a common astrometric reference frame; and
iii. combine astrometric and photometric measurements for all images within an epoch to estimate the average positions, fluxes, and associated errors.

The final product is a catalog for each epoch and filter of stellar fluxes in instrumental magnitudes and positions in pixels in a camera coordinate system. Instrumental fluxes are converted to Vega magnitudes as described in Section 3.2. Each step is described in more details below.

First, stellar fluxes and positions are initially extracted from the individual flt images using point-spread function (PSF)-fitting methods and the `HST1PASS` software adapted from J. Anderson (2022). During the source extraction process, the known camera distortions are corrected for both ACS-WFC and WFC3-IR[12] (J. Anderson & I. R. King 2006).

We derive the first-order coordinate transformations of all the star lists with `HST1PASS`, based on the 2005 ACS-WFC F814W data as an initial reference frame, which has the finest resolution. Reference stars are chosen to fall within 0.07 ACS pixels or 3.5 mas and have an instrumental F814W magnitude between −13.5 and −10.0. Uncertainties on the positions and fluxes are derived from the rms error of the measurements in the individual exposures and are typically below 0.5 mas for the brightest stars. This process is repeated using the averaged star list from the first pass as a new reference frame and using fainter stars in the transformation. We note that the 2005 ACS-WFC F814W star list is also transformed to treat the two chips as independent images, each with its own transformation. This second pass reduces uncertainties by a factor of 4 to 0.006 ACS pixels, or 0.3 mas.

Preliminary PMs are determined by identifying cluster members, which are then used to establish a refined set of reference stars. The reference stars are selected based on the following properties:

i. preliminary PMs within 0.7 mas yr$^{-1}$ of the cluster's mean motion;
ii. preliminary PM uncertainties less than 0.2 mas yr$^{-1}$; and
iii. photometric uncertainties of less than 5% in F814W and 10% in the IR passbands.

The entire process of transforming individual exposures is repeated with the new reference frame. The final transformation residuals are listed in Table 1 for each epoch and filter.

With the image coordinate transformations in hand, we use the sophisticated source detection routine `ks2` (J. Anderson & I. R. King 2006; J. Anderson et al. 2008; A. Bellini et al. 2017, 2018) to extract stellar fluxes and positions from a stack of images, which results in a much deeper catalog. The positions and fluxes of these stars are measured from the individual images using the individual-image PSFs described earlier.

---
[12] The WFC3-IR distortion solution can be downloaded from https://www.stsci.edu/files/live/sites/www/files/home/hst/instrumentation/wfc3/data-analysis/psf/_documents/STDGDC_WFC3IR.fits.

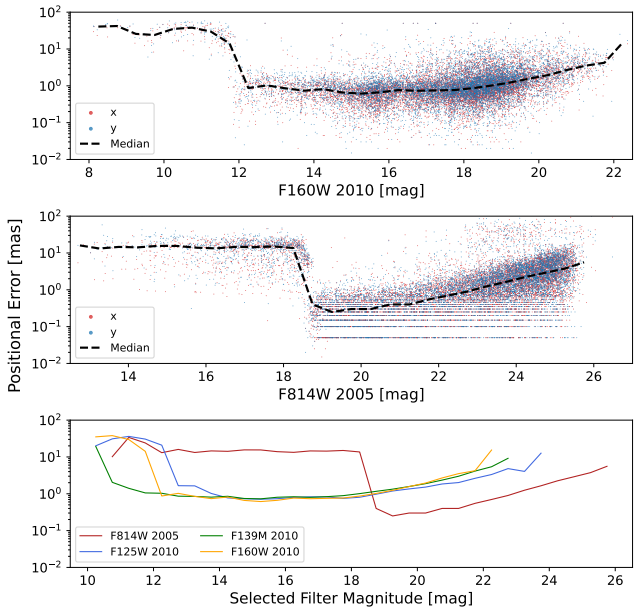

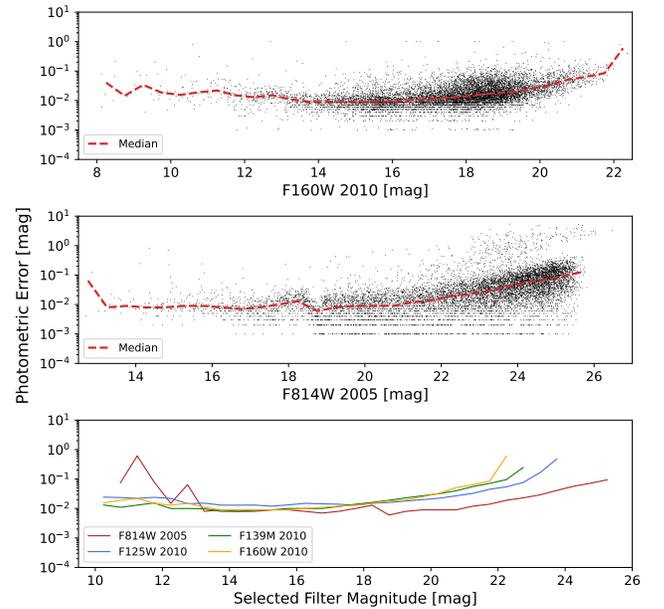

**Figure 2.** Astrometric uncertainty vs. brightness for all filters and years. Top: positional uncertainty in the F160W 2010 epoch. Middle: Positional uncertainty in the F814W 2005 epoch. Bottom: median uncertainties for each epoch. The astrometric uncertainty is the rms error over the individual exposures in each filter and epoch. The uncertainties in *x* (red) and *y* (blue) are oriented along the detector coordinates of the 2005 F814W observations. The dashed black lines in the first two panels show the median error after rejecting outliers larger than $3\sigma$.

**Figure 3.** Photometric uncertainty vs. brightness for selected filters and years. Top: photometric uncertainty in F160W 2010 epoch. Middle: photometric uncertainty in F814W 2005 epoch. Bottom: median photometric uncertainty in each epoch. The photometric uncertainty is the rms error over the individual exposures in each filter and year. The dashed red lines in the first two panels show the median error after rejecting large ($>3\sigma$) outliers.

**Table 2**
Photometric Zero-points

| Filter | Zero-points (mag) | |
|---|---|---|
| | Aperture[a] | ks2 |
| F814W | 25.518 | 32.678 $\pm$ 0.010 |
| F125W | 25.144 | 25.231 $\pm$ 0.010 |
| F139M | 23.209 | 23.284 $\pm$ 0.010 |
| F160W | 24.504 | 24.570 $\pm$ 0.010 |

**Note.**
[a] 2005 zero-point value for F814W and 2012 photometric calibration values for the $R = 0.4$ aperture for F125W, F139W, and F160W.

The positions of stars from each exposure are averaged within each epoch, and the uncertainties of the positions and fluxes are estimated in the same manner as described above. Figures 2 and 3 show the final astrometric and photometric rms errors, respectively, for all stars in the catalog.

### 3.2. Photometric Calibration

The final catalogs are photometrically recalibrated since the published zero-points for ACS-WFC and WFC3-IR are derived from aperture photometry rather than PSF fitting. We calculate a new photometric zero-point for each filter with the following procedure. First, we download the drizzle-combined mosaics from the HST archive and perform aperture photometry on the images using an aperture radius of 0″.4 for the WFC3-IR images and 0″.5 for the ACS images. The apparent magnitudes for the stars in the drizzled images were determined using the Space Telescope Science Institute's published 2012 photometric zero-points with an $R = 0.4$ aperture for the WFC3-IR filters,[13] and the 2005 zero-point for F814W.[14]

We crossmatch stars between our final catalogs from ks2 and the drizzled-image star lists. A set of photometric calibration stars is selected to be bright but not saturated in the instrumental magnitude range of $-11.0$ to $-9.5$ and to be isolated with no neighbors of comparable magnitudes within 0.25 mas. These stars are used to derive the average flux ratio and new zero-points for our PSF photometry. Table 2 contains the zero-points for both the 2012 photometric calibration with the $R = 0.4$ aperture and ks2 photometry for all filters.

---
[13] https://www.stsci.edu/hst/instrumentation/wfc3/data-analysis/photometric-calibration/ir-photometric-calibration
[14] https://acszeropoints.stsci.edu

### 3.3. Final Proper Motions

PMs are derived with the linear fit to astrometric positions for stars detected in all four epochs, including 2005 F814W, 2010 F160W, 2013 F160W, and 2015 F160W observations:

$$\alpha^* = \mu_{\alpha^*}(t - t_0) + \alpha_0^*$$
$$\delta = \mu_\delta(t - t_0) + \delta_0, \quad (1)$$

where $t_0$ is the average time weighted by the inverse square of the astrometric uncertainties, $(\alpha^*, \delta)$ is the observed position at time $t$, and $(\alpha_0^*, \delta_0)$ is the position at reference time $t_0$. We adopt the flattened R.A. $\alpha^* = \alpha \cos\delta$ in this study. PMs are fit to the positions weighted by the inverse square of the positional error in each epoch (see A. M. Ghez et al. 2005; J. R. Lu et al. 2009; S. Yelda et al. 2014; M. W. Hosek et al. 2015, for a more complete description). The resulting observed catalog contains 10,346 stars with PMs and associated uncertainties.

We imposed a conservative constraints on PMs, selecting stars within PMs within $3\sigma$ of the mean values in both the *x*-

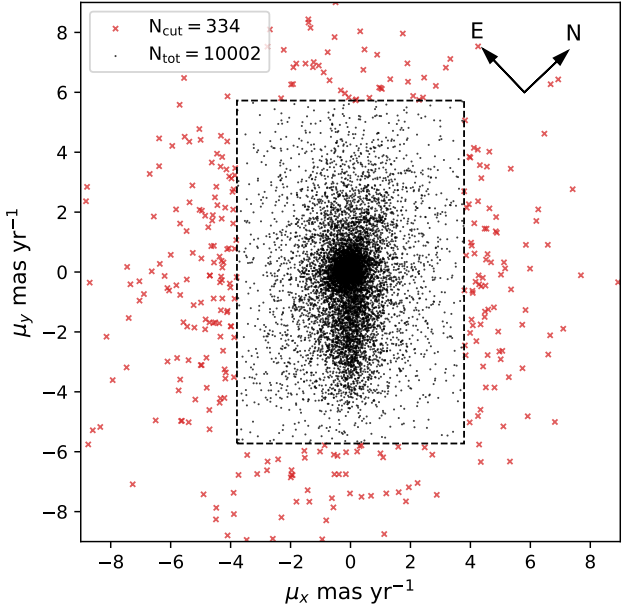

**Figure 4.** PM vector point diagram (VPD) of the entire catalog. The central box outlined by the black dashed lines shows the $3\sigma$ cut in both directions. The kept stars are marked by black points inside the box, and the cut stars are marked by red crosses outside the box.

and y-directions. The resulting constraint is

$$|\mu_x - \bar{\mu}_x| < 3.79 \text{ mas yr}^{-1},$$
$$|\mu_y - \bar{\mu}_y| < 5.76 \text{ mas yr}^{-1}. \quad (2)$$

At the best-fit distance of 3.7 kpc, which we will discuss in Section 5.3, they correspond to a velocity of 66.94 km s$^{-1}$ and 101.66 km s$^{-1}$ in the x- and y-direction, respectively. This cut is visualized in Figure 4 in velocity space. We instituted the box cut to ensure that our membership selection process is not hampered by stars at the extreme edges of the PM distribution. This cut removes 344 stars from the catalog. The catalog contains 10,002 stars for the PM membership analysis.

## 4. Methods: Modeling the Cluster

### 4.1. Cluster Membership

We distinguish cluster members from contaminating field stars by analyzing the PM distribution and the CMD. Each star in the catalog is assigned a PM membership probability, $p_\mu$, and a color membership probability according to its location in the CMD space, $p_{\text{color}}$. The PM membership $p_\mu$ is a continuous variable that ranges from zero to one, whereas $p_{\text{color}}$ is a Boolean criterion. The final membership value is the product of the two components, $p_{\text{clust}} = p_\mu \cdot p_{\text{color}}$. Only stars with $p_{\text{clust}} \geqslant 0.3$ are included in the structure and radial profile analysis.

First, we determine the PM membership. As cluster members tend to move with a systemic PM, they can be distinguished from the field populations kinematically and should form a compact region on the VPD, a diagram of the PM vectors in the x- and y-direction, respectively. To robustly identify the cluster members in the region, we adopt a Gaussian mixture model (GMM) to model the PM distribution of cluster and field stars with Bayesian inference. This model allows us to assign a kinematic cluster membership probability to each star based on its PM. The model employs a mixture of N Gaussians to represent the cluster and field star populations (see W. I. Clarkson et al. 2012; M. W. Hosek et al. 2015; N. Z. Rui et al. 2019, for details), with N ranging from two to five explored in this analysis. We define the likelihood function for the set of N measured stars, $\mathcal{L}$:

$$\mathcal{L} = \prod_i^N L(\mu_i), \quad (3)$$

where $L(\mu_i)$ is the likelihood of the ith star with a measured PM vector $\boldsymbol{\mu}_i \equiv (\mu_{\alpha*}, \mu_\delta)$, defined as

$$L(\mu_i) = \sum_{k=0}^{K} \frac{\pi_k}{2\pi |\Sigma_{ki}|^{\frac{1}{2}}}$$
$$\times \exp\left[-\frac{1}{2}(\boldsymbol{\mu}_i - \bar{\boldsymbol{\mu}}_k)^\top \Sigma_{ki}^{-1}(\boldsymbol{\mu}_i - \bar{\boldsymbol{\mu}}_k)\right], \quad (4)$$

for K Gaussian components, where $\pi_k$ is the fraction of stars in the kth Gaussian, $\bar{\boldsymbol{\mu}}_k$ is the PM centroid vector of the kth Gaussian, and $\Sigma_{ki}$ is the covariance matrix of the kth Gaussian and the ith star with PM measurements of $\boldsymbol{\mu}_i$ and an associated uncertainty of $\boldsymbol{\epsilon}_i \equiv (\epsilon_{\alpha*,i}, \epsilon_{\delta,i})$. Following M. W. Hosek et al. (2015) and W. I. Clarkson et al. (2012) we take

$$\Sigma_{ki} = S_i + Z_k, \quad (5)$$

where $S_i$ is the diagonal component of the velocity error matrix:

$$S_i = \begin{bmatrix} \epsilon_{\alpha*,i}^2 & 0 \\ 0 & \epsilon_{\delta,i}^2 \end{bmatrix}, \quad (6)$$

and $Z_k$ is the covariance matrix of the kth Gaussian component:

$$Z_k = \begin{bmatrix} \sigma_{\alpha*}^2 & \rho\sigma_{\alpha*}\sigma_\delta \\ \rho\sigma_{\alpha*}\sigma_\delta & \sigma_\delta^2 \end{bmatrix}. \quad (7)$$

Here $\sigma_{\alpha*,i}$ and $\sigma_{\delta,i}$ denote the intrinsic velocity dispersion in the R.A. converted by $\cos\delta$ and the decl. direction, respectively, and $\rho$ denotes the correlation coefficient between the two components.

With the likelihood function, the posterior probability distribution is determined by Bayes' theorem:

$$P(\boldsymbol{\pi}, \bar{\boldsymbol{\mu}}, \boldsymbol{Z} | \boldsymbol{\mu}, \boldsymbol{S}) = \frac{P(\boldsymbol{\mu}, \boldsymbol{S} | \boldsymbol{\pi}, \bar{\boldsymbol{\mu}}, \boldsymbol{Z}) P(\boldsymbol{\pi}, \bar{\boldsymbol{\mu}}, \boldsymbol{Z})}{P(\boldsymbol{\mu}, \boldsymbol{S})}, \quad (8)$$

where $P(\boldsymbol{\pi}, \bar{\boldsymbol{\mu}}, \boldsymbol{Z} | \boldsymbol{\mu}, \boldsymbol{S})$ is the posterior probability of the model, $P(\boldsymbol{\mu}, \boldsymbol{S} | \boldsymbol{\pi}, \bar{\boldsymbol{\mu}}, \boldsymbol{Z})$ is the probability of the observed velocity distribution given the model, $P(\boldsymbol{\pi}, \bar{\boldsymbol{\mu}}, \boldsymbol{Z})$ is the prior probability of the model, and $P(\boldsymbol{\mu}, \boldsymbol{S})$ is the sample evidence. Here, $\boldsymbol{\pi}$ is the set of $\pi_k$ values, $\bar{\boldsymbol{\mu}}$ is the set of Gaussian velocity centroids, $\boldsymbol{Z}$ is the set of Gaussian covariance matrices, $\boldsymbol{\mu}$ is the set of observed stellar PMs, and $\boldsymbol{S}$ is the set of PM error matrices.

To fit the GMM, we use MultiNest (F. Feroz et al. 2009), a multimodal nested sampling algorithm, and its Python wrapper, PyMultiNest (J. Buchner et al. 2014). To determine the merit of each K Gaussian model, we compare the results of their Bayesian information criterion (BIC) tests (G. Schwarz 1978). The BIC regularizes a model by modifying the fit residuals with a penalty for model complexity. We find

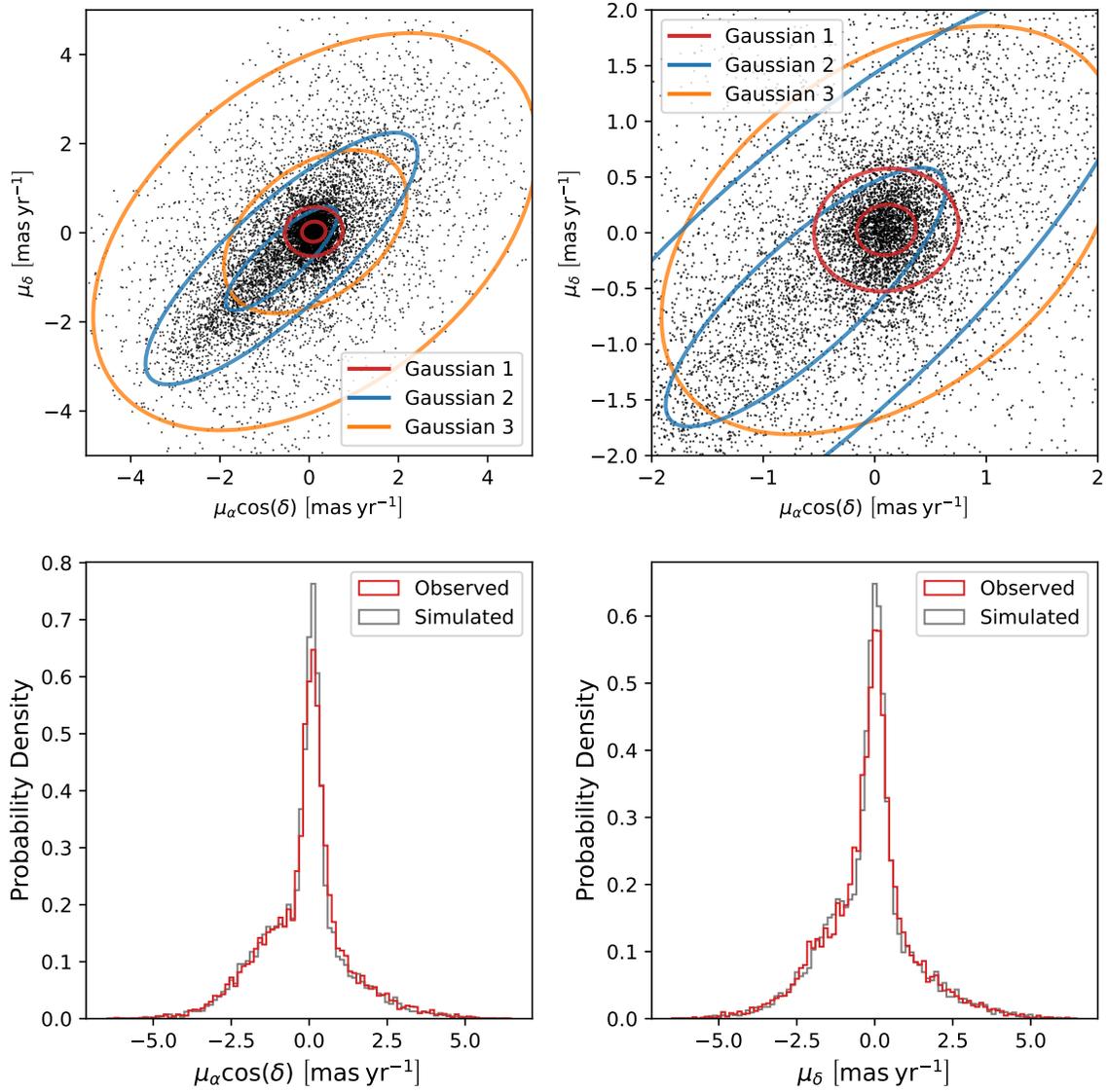

**Figure 5.** GMM fit and probability densities of the PMs. Top: three-component GMM fit of the PMs shown in the top left panel and a zoomed-in view of the top right panel. Each Gaussian model is shown in its $1\sigma$ and $3\sigma$ iso-density ellipses. The first Gaussian depicted in red models the probability of being a cluster member, while the second and third Gaussians, shown in blue and amber, respectively, model the field star probabilities. Bottom: probability density distributions of PMs in the R.A. direction shown in the bottom left panel and decl. direction in the bottom right panel. Observed stars and simulated stars, assuming the GMM perfectly describes their PM, are shown in red and gray, respectively.

Table 3
Kinematic Membership Gaussian Mixture Model: Parameters, Priors, and Results

| Parameters | Unit | Cluster Gaussian Prior[a] | Cluster Gaussian Result | Field Gaussian 1 Prior | Field Gaussian 1 Result | Field Gaussian 2 Prior | Field Gaussian 2 Result |
|---|---|---|---|---|---|---|---|
| $\pi_k$ | ⋯ | $U(0, 1)$ | $0.33 \pm 0.01$ | $U(0, 1)$ | $0.34 \pm 0.01$ | $U(0, 1)$ | $0.33 \pm 0.01$ |
| $\mu_{\alpha*,k}$ | mas yr$^{-1}$ | $\mathcal{N}(0, 0.3)$ | $-0.05 \pm 0.01$ | $U(-10, 10)$ | $0.01 \pm 0.02$ | $U(-10, 10)$ | $-0.08 \pm 0.04$ |
| $\mu_{\delta,k}$ | mas yr$^{-1}$ | $\mathcal{N}(0, 0.3)$ | $0.09 \pm 0.01$ | $U(-10, 10)$ | $-0.85 \pm 0.07$ | $U(-10, 10)$ | $0.11 \pm 0.08$ |
| $\sigma_k$ | mas yr$^{-1}$ | $U(0, 1)$ | $0.27 \pm 0.01$ | $U(0, 8)$ | $1.64 \pm 0.05$ | $U(0, 8)$ | $2.33 \pm 0.05$ |
| $\epsilon_k$ | ⋯ | $U(0, 1)$ | $0.84 \pm 0.04$ | $U(0, 1)$ | $0.30 \pm 0.02$ | $U(0, 1)$ | $0.62 \pm 0.02$ |
| $\theta_k$ | rad | $U(0, \pi)$ | $0.99 \pm 0.15$ | $U(0, \pi)$ | $1.55 \pm 0.02$ | $U(0, \pi)$ | $1.47 \pm 0.03$ |

**Notes.** Parameter description: $\pi_k$: normalized fraction of the $k$th Gaussian. $\mu_{\alpha*,k}$, $\mu_{\delta,k}$: mean PM of the $k$th Gaussian in $\alpha^*$ and $\delta$. $\sigma_k$: standard deviation of the $k$th Gaussian in the semimajor axis direction. $\epsilon_k$: minor-to-major axis ratio, or ellipticity of the $k$th Gaussian. $\theta_k$: rotation angle of the semimajor axis of the Gaussian ellipse from the positive $x$-direction.
[a] $U(a, b)$ stands for a uniform distribution between $a$ and $b$, and $\mathcal{N}(\mu, \sigma)$ denotes a normal distribution with a mean of $\mu$ and a standard deviation of $\sigma$.

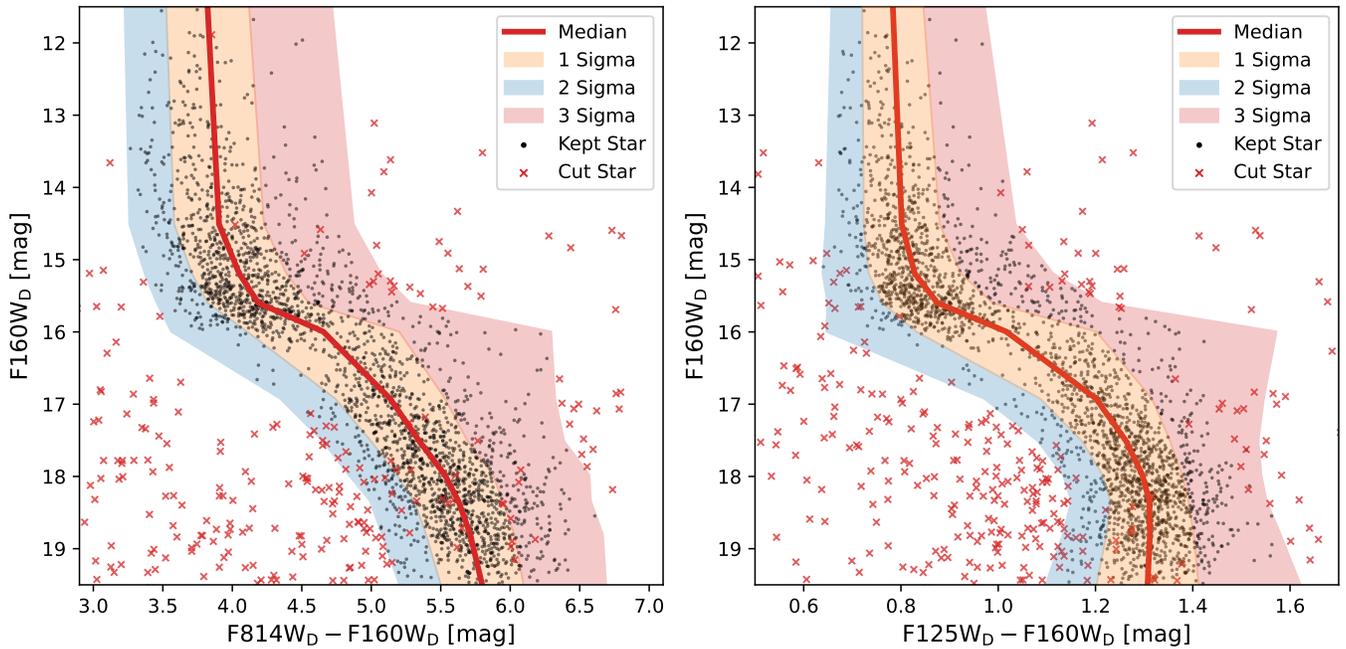

**Figure 6.** Color membership determination CMDs for F814W − F160W (left) and F125 − F160W (right). Stars with $p_\mu \geqslant 0.7$ are used to determine the median and standard deviation in the CMD space. The yellow-shaded region represents $1\sigma$ from the median. The blue and red shaded regions correspond to stars less than $2\sigma$ bluer and $3\sigma$ redder than the median, respectively. Stars $2\sigma$ bluer or $3\sigma$ redder than the median in each magnitude bin, or, equivalently, those outside any shaded regions, are excluded from our final analysis. Stars marked by red crosses are the cut stars, while black dots are the kept ones. Here, the bins have been adaptively sized to contain approximately 287 stars in each of the 10 bins. Contaminating objects within our restricted cluster sequence may be binaries or field stars.

the $N = 3$ model achieves the balance between describing the data well and the complexity of the model itself. The parameters, priors, and results are summarized in Table 3, and the Gaussians on the VPD are shown in Figure 5. We generate a simulated set of stars assuming the modeled GMM perfectly describes the PMs, and the comparison between the simulated and observed PMs is shown in the bottom panels in Figure 5. Note that the GMM modeling of the VPD is consistent across magnitudes, and we find no benefit in performing magnitude-by-magnitude GMM modeling with this data set.

With the GMM model, the PM membership probability $p_\mu^i$ is determined as

$$p_\mu^i = \frac{\pi_1 P_1^i}{\sum_{k=1}^{3} \pi_k P_k^i},\qquad(9)$$

where $\pi_k$ is the normalized fraction of the $k$th Gaussian, and $P_k^i$ is the probability of the $i$th star being in the $k$th Gaussian.

Wd1 is known to be located within the Scutum–Crux Arm, and kinematic confusion in the region can be a source of contamination (I. Negueruela et al. 2022). However, with unprecedented PM measurements and our methodology of assigning membership probabilities, instead of a decisive determination of members and nonmembers, we expect this effect to be limited. Additionally, the directional preference for the PMs of the field stars indicated by the highly eccentric ellipsoids of the second and third Gaussians is likely a result of the field stellar population in the Scutum–Crux Arm.

Next, we determine the color membership. Since cluster members are assumed to have the same age, distance, and metallicity, they are expected to follow a distinct sequence in CMD space. Thus, we can eliminate stars with photometry inconsistent with the cluster sequence as likely field contaminants. Applying color cuts thus further reduces field contamination. After masking the low-completeness region, we select stars with high kinematic membership probabilities with $p_\mu \geqslant 0.7$, and determine the median and standard deviation of colors in differentially dereddened magnitude bins, each containing 287 stars. Note that this stricter kinematic membership is only utilized for determining the color membership. For the structural and kinematic analysis, we adopt the $p_\mu \geqslant 0.3$ criterion. We will illustrate the dereddening and completeness in Sections 4.2 and 4.3, respectively. Figure 6 shows the CMD with our $1\sigma$, $2\sigma$, and $3\sigma$ masks in F160W$_D$ versus F814W$_D$ − F160W$_D$ and F160W$_D$ versus F125W$_D$ − F160W$_D$ space. Stars bluer than $2\sigma$ or redder than $3\sigma$ than the bulk cluster in either CMD space are assigned $p_{\rm color} = 0$, and otherwise $p_{\rm color} = 1$. We adopted a stricter cut on the blue side because some fraction of cluster members might be expected to have extra intrinsic reddening from circumstellar disks, given the youth of the cluster.

The final membership $p_{\rm clust}$ is then the product of PM membership and color membership:

$$p_{\rm clust} = p_\mu \times p_{\rm color}.\qquad(10)$$

We require $p_{\rm clust} \geqslant 0.3$ to be used in our analysis, resulting in a catalog of 3586 unweighted stars. Histograms of the PM membership and cluster membership probabilities are shown in Figure 7 in blue and red, respectively. Note that this is not the final CMD we use for the structural and dynamic analysis. Rather, this only serves as an illustration of the color membership determination criterion, in which we utilized the stricter $p_\mu \geqslant 0.7$ kinematic membership criterion to obtain a clean CMD. We incorporate stars with $p_{\rm clust} \geqslant 0.3$ for the analysis.

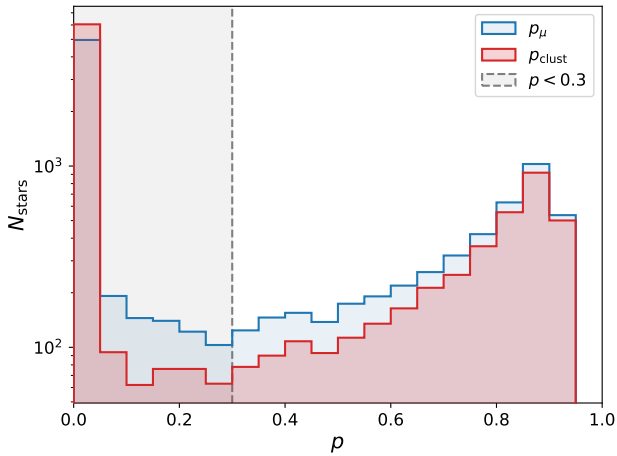

Figure 7. Histogram of PM membership and cluster membership probability. The blue and red histograms show the PM membership $p_\mu$ and the cluster membership probability $p_{\rm clust} = p_\mu \cdot p_{\rm color}$, respectively. The gray dashed line and the shaded region mark the nonclusters members with $p_{\rm clust} < 0.3$.

### 4.2. Extinction

As the cluster is subject to known reddening (I. Negueruela et al. 2022), we correct for extinction by producing a spatial attenuation map and differentially deredden the stars in our sample. In the following analysis, we use the subscript "D" after the filter names to denote a dereddened magnitude. We create an extinction map for the field based on individual extinction values derived from high-probability MS cluster members. The intrinsic colors of such stars are nearly independent of mass in the near-IR (NIR) filters. Therefore, their extinction can be estimated from their observed color, given an assumed distance and extinction law.

To determine the attenuation value $A_{Ks}$ of the MS stars, we used SPISEA (M. W. Hosek et al. 2020) to produce a reference isochrone with a reference distance of 4000 pc, age of 7 Myr, and solar metallicity. The age and distance are adopted from literature values (e.g., E. R. Beasor et al. 2021; F. Navarete et al. 2022) as described in Section 1. Wd1 is known to have a marginally supersolar metallicity. For instance, N. Yusof et al. (2022) reproduce the observed populations of W-R stars and supergiants in Wd1 with a $Z = 0.02$ model, or $[Z] = \log_{10} Z/Z_\odot = 0.17$. However, the merged stellar model in the analysis software SPISEA only has solar metallicity currently, and is still being developed to incorporate the latest updates. We will test the supersolar metallicity using MIST stellar models and show that the effect is negligible in the scope of this work.

We emphasize that this isochrone is only used for reference in calculating the extinction, as the colors of MS stars used to derive the extinction values are insensitive to the cluster age and distance within a reasonable range. We present the modeling of the age and distance of Wd1 in CMD space in Section 4.5. The reference isochrone adopted a mixed set of stellar evolutionary models, including the Baraffe (I. Baraffe et al. 2015), Pisa (E. Tognelli et al. 2011), and Geneva (S. Ekström et al. 2012) models.[15] The isochrone also utilizes a mixed atmospheric model consisting of ATLAS9 (F. Castelli & R. L. Kurucz 2003), PHOENIX version 16 (T. O. Husser et al. 2013), and BT-Settl models (F. Allard et al.

---

[15] spisea.evolution.MergedBaraffePisaEkstromParsec

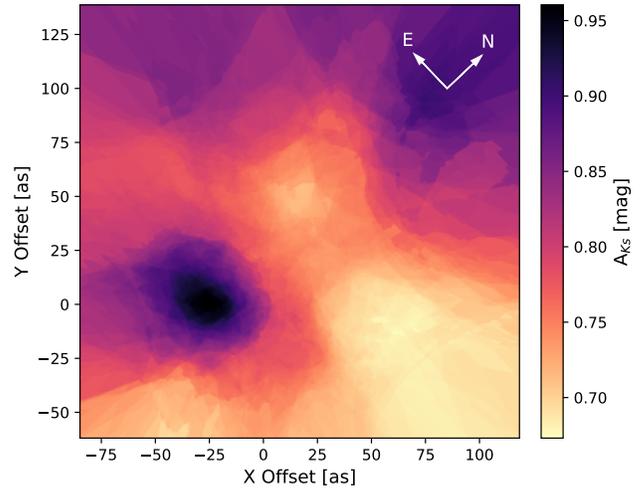

Figure 8. IR extinction map produced by analysis of the extinction values of the MS stars. The map axes are referenced to the non-completeness-corrected centroid of the cluster, weighted by the dereddened F160W$_D$ magnitudes. The assigned pixel values are taken from the mean of the 30 nearest $3\sigma$ clipped stars.

2012a, 2012b; I. Baraffe et al. 2015).[16] We refer the readers to M. W. Hosek et al. (2020) for detailed information.

We then varied the total extinction of the reference isochrone to measure the reddening vector of the MS stars, which describes the shift of a star's position in the CMD with changing extinction. We adopt the revised extinction law of M. W. Hosek et al. (2019), which was derived using stars from Arches and Wd1 for optical and NIR extinction in highly reddened regions. We generate two reference isochrones, one at $A_{Ks} = 0.73$ mag and another at $A_{Ks} = 1.03$ mag, to measure the reddening vector slope, as these values cover most of the MS stars.

We produced a pixel-by-pixel reddening map by utilizing high kinematic membership probability ($p_{\rm clust} \geqslant 0.7$) MS stars brighter than F160W $= 15.0$ mag, which is 0.5 mag brighter than the PMS turn-on. We interpolate along mass in each isochrone and map reddening vectors as a function of mass between the fiducial and secondary isochrones. Each star is assigned an extinction value based on the distance from the reference isochrone along the nearest reddening vector. We reject stars with $A_{Ks}$ values further than $3\sigma$ away from the mean $A_{Ks}$, as well as stars with a distance further than 0.25 mag from the nearest reddening vector on the CMD. After the rejection, we are left with 451 stars out of 2934 for assigning extinction values, with a median $A_{Ks} = 0.79 \pm 0.06$ mag.

Each pixel is assigned an extinction value and error via the $3\sigma$ clipped mean of the 30 nearest neighbors among the 451 stars used to calculate extinction. The resulting IR extinction map is shown in Figure 8. The extinction error map ($\sigma_{\rm map}$) is calculated as $\sigma_{A_{Ks}} / \sqrt{N}$, where $\sigma_{A_{Ks}}$ is the standard deviation of the extinction values and $N$ is the number of stars used. Stars in the catalog are assigned the extinction and error of the pixel they fall on. We build separate extinction maps from the optical and IR photometry to see which one results in a tighter cluster on the CMD.

In this work, we adopted the IR extinction map in F125W, shown in Figure 8. We noticed that the optical and IR extinction maps are remarkably consistent with the assumed extinction law, suggesting that the adopted extinction law is a good match for the data. Both maps show significant spatial variability over a similar range of extinction values, with $\Delta A_{Ks} = 0.29$ mag for both the optical and IR CMD-based maps. The median of both maps is $A_{Ks} = 0.79 \pm 0.06$ mag, with the error being the standard deviation of the map. The maps can differ by as much as $|A_{Ks}| = 0.03$ mag with a median absolute difference of $A_{Ks} = 0.004$ mag. In addition, the median pixel value errors are both $A_{Ks} = 0.013$ mag for the optical and IR CMD-based maps. The maximum absolute difference between the pixel value errors of the two maps is 0.012 mag.

---

[16] spisea.atmospheres.get_merged_atmosphere

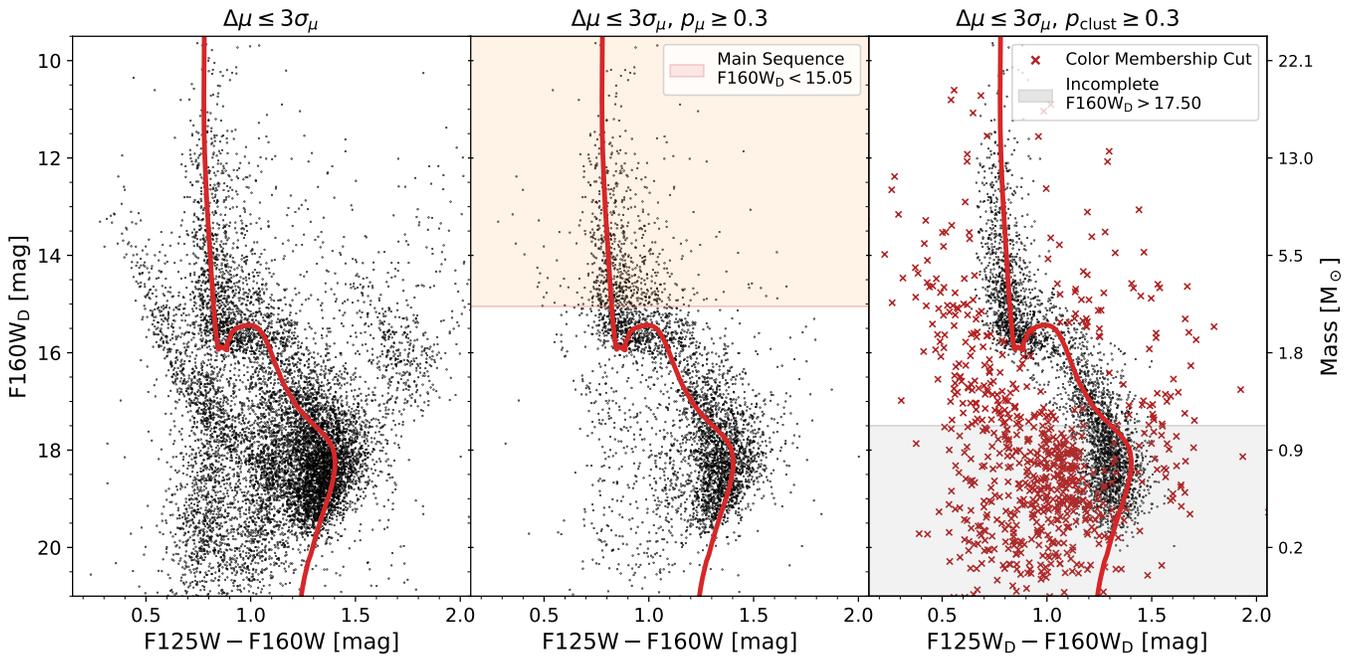

**Figure 9.** Left: CMD of the cluster catalog after applying a $3\sigma_\mu$ box cut in the *x*- and *y*-directions. Center: CMD of cluster PM membership $p_\mu \geq 0.3$ and reference isochrone with an $A_{Ks}$ of 0.73 mag, distance of 4000 pc, and age of 7 Myr. The orange shaded region indicates the faintest dereddened magnitude limit of $F160W_D = 15.05$ of the stars used to produce the extinction map. Right: CMD of cluster members with $p_{clust} \geq 0.3$ after extinction correction and color membership cut, with rejected stars marked with red crosses. The shaded gray region corresponds to stars fainter than our completeness cut of 17.5 mag.

Our extinction maps are consistent with the observations of W. Brandner et al. (2008), in which the authors claimed that the regions east and north of the cluster center tend to have slightly higher extinctions, while the regions west and south of the cluster have slightly lower extinctions. B. Lim et al. (2013) also found the east side of the field to have higher extinction than the west side, though they also observed the highest reddening toward the cluster center. In our extinction maps, a relatively high patch of reddening is found to the southeast of the uncorrected cluster center. The cause of this discrepancy is likely due to the smaller sample size of B. Lim et al. (2013), who measured the reddening of 53 OB supergiants. Our reddening sample is much larger, allowing us to map out the differential reddening in more detail.

Figure 9 shows the CMDs before and after differentially dereddening the colors, along with the color membership cut. The magnitude limit of $F160W_D = 15.05$ mag for deriving the extinction map is shown as the orange shaded region in the middle panel. The red curves indicate the best-fit isochrone rather than the reference isochrone. The determination of the cluster's age and distance to generate the best-fit isochrone will be elaborated in Section 4.5.

Notice that the dereddened color of the stars fainter than $F160W_D > 17.5$ differs from that of the best-fit isochrone as shown in Figure 9, with the observed stars bluer than the theoretical isochrone. This is a caveat arising from the poor fit in the PMS regime of the CMD. Adopting the alternative reddening law presented in A. Damineli et al. (2016), which is specifically derived from Wd1, has little impact in resolving the discrepancy. The maximum difference in the $F125W_D - F160W_D$ color is as little as 0.03 mag, and the mean difference is less than 0.01 mag. In comparison, the observed color discrepancy between the stars and the isochrone is about 0.1 mag. We also examined whether the discrepancy could be attributed to a difference in the reddening vector slope for the PMS stars, since the reddening vector may vary with intrinsic star color. However, with the SPISEA isochrones, we found that the change in the dereddened $F125W_D - F160W_D$ colors of the PMS stars is $<0.03$ mag due to this effect, insufficient to explain the color difference. Additionally, we investigated the effect of metallicity and tested the MIST model from $[Z] = 0 - 0.3$. The maximum color change in $F125_W - F160_W$ is 0.04 mag at the dimmest end of the isochrone at $F160W = 21$ mag. Again, this is insufficient to explain the deviation alone. Theoretical stellar evolutionary models are rarely tested comprehensively across such a wide range of stellar masses with completeness correction within a single environment (e.g., L. Haemmerlé et al. 2019), and PMS stars are known to be problematic. For instance, the same color discrepancy trend in Wd1 can be seen in the clean CMD in Figure 5 in M. Gennaro et al. (2011). The fainter stars in Pleiades are also observed to deviate from the MIST model in the CMD space, as shown in the middle panel of Figure 26 in J. Choi et al. (2016). With the detailed structural modeling of Wd1 presented in this work, the cluster emerges as a promising and comprehensive test bed for stellar evolutionary models. A combination of the aforementioned factors could contribute to the mismatch of the isochrones. In a forthcoming study, we will explore various stellar models and see if they help reconcile the discrepancy observed in this work. We emphasize that our analysis is largely unaffected by this discrepancy, as it will be restricted to stars brighter than F160W$_D \leqslant 17.5$ set by the completeness limit, which will be discussed in detail in Section 4.3.

### 4.3. Completeness

As the region is crowded, completeness correction is vital to account for unobserved stars contaminated by bright stars. We estimate the completeness of our final catalog by planting a set of artificial stars into the original images and processing them through our entire analysis pipeline. The procedure for this analysis has been described in detail in earlier work (J. R. Lu et al. 2013; M. W. Hosek et al. 2015, 2019; N. Z. Rui et al. 2019) and entails planting and extracting 600,000 artificial stars. The positions of the artificial stars are randomly sampled from a uniform distribution over the FOV. The fluxes of the artificial stars are generated to thoroughly cover the color–magnitude space populated by the observed stars with some additional padding on all sides.

We modify the astrometric and photometric errors of the detected artificial stars to incorporate additional systematic uncertainties, potentially arising from PSF variability, intra-pixel sensitivity variations, and uncorrected residual distortions in the simulations (M. W. Hosek et al. 2015). The systematic error terms are then added in quadrature, ensuring that the resulting error distributions at different magnitudes for the simulated stars match those of the observed stars.

Artificial stars are matched across all filters and epochs, and their PMs are fit in the same manner as the observed data. After accounting for systematic errors in the observed stars, the final distribution of PM errors for the artificial stars matches the observed stars. There are 423,790 out of 600,000 stars detected in all four epochs: 2005 F814W, 2010 F160W, 2013 F160W, and 2015 F160W. This catalog is used to construct a spatial completeness map and completeness curves as a function of magnitude.

To construct the completeness map, we first produce a grid of reference points one by one arcsecond in the $x$- and $y$-directions. We determine the completeness of each reference point from its nearest 2000 neighbors in each filter and each magnitude bin ranging from 8 to 32 mag with a step size of 0.5 mag. Each reference pixel is assigned a completeness value for each magnitude bin. Observed stars are then assigned the completeness value of the nearest reference point.

We masked the regions affected by low completeness smaller than 15% at F160W$_D$ = 17.5, or equivalently 1.12 $M_\odot$, as shown in the white contours in Figure 10. We tested the completeness correction in the unmasked FOV and in the lowest completeness region marked by the cyan ellipse with an effective radius of 0.35 pc, ensuring the completeness levels are reasonable for the faintest magnitude. The size, orientation, and elongation of the ellipse are derived from the 2D Gaussian profile fit to the stellar density map, which will be described in Section 4.4. The effective radius accounts for the elongation and orientation of the cluster and will be defined in that section.

To check the completeness correction, we first construct 1D completeness curves by interpolating along the magnitudes for all of the remaining fields and the central low-completeness region, respectively, as shown in Figure 11. As can be seen from the completeness curves in the right panel in Figure 11, cutting stars fainter than F160W$_D$ = 17.5 is equivalent to applying a 37% completeness limit in the lowest completeness region.

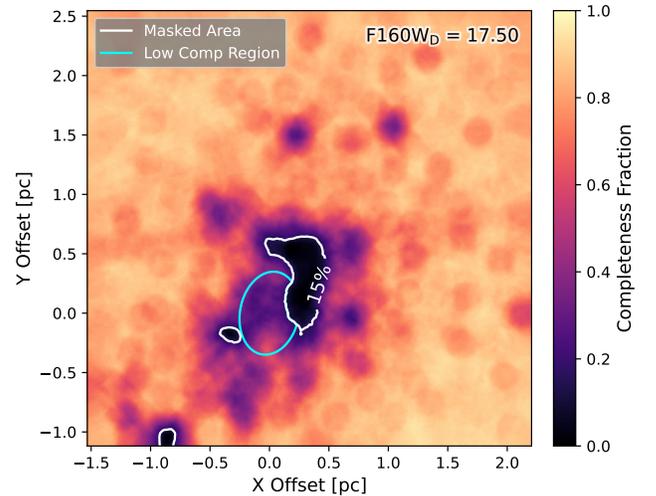

**Figure 10.** Two-dimensional completeness map for dereddened F160W$_D$ mag of 17.5. The white contours show the 15% low-completeness regions, which are masked in the radial profile analysis. The partial cyan ellips*e* shows the innermost radial bin for completeness check with an effective radius of 0.29 pc.

Furthermore, we check the independence of the field stellar density on the effective radius, as is expected for the small FOV of our observations. For this purpose, we interpolate the completeness for each star only as a function of the effective radius and magnitude, instead of using the pixel completeness calculated as in Figure 10. Note that the interpolated completeness is only used to check the field stellar density, and the pixel completeness is used for the actual analysis. We use 2D interpolation on a magnitude grid ranging from 10 to 22 mag in F160W$_D$ with a step size of 0.5 mag, and on the 10 radial bins, which equally partition the corrected stellar count. The resulting cluster and field stellar densities as a function of effective radius after correcting for membership, completeness, and area fraction are shown in Figure 12. To confirm the independence of the field stellar density on effective radius after completeness correction, we performed a Kolmogorov–Smirnov (K-S) test on 10,000 simulated field stellar density arrays of length 10, same as the length of radial bins in Figure 12, with each element following a normal distribution centered at the measured value and a standard deviation being the associated uncertainty in that bin. The simulated data are tested against a normal distribution with the expectation being the mean of the measured field stellar density across the effective radius, and the standard deviation being the standard deviation of 10 measured field stellar densities. The null hypothesis is that the field stellar density in each bin follows the same Gaussian distribution, independent of the effective radius. The resulting $p$-value of the 10,000 simulations is $0.50 \pm 0.27$, significantly greater than the critical value of 0.05. Therefore, we accept the null hypothesis that the field stellar density is independent of effective radius, proving the validity of our completeness correction.

We note that an alternative approach to the completeness correction was implemented to account for possible nonmodeled magnitude-dependent PM errors that might be sufficiently large to alter the assignment of stars to field versus Wd1, and change the geometric pattern of the GMM components. For this, we first stratified the observed stars into 1 mag bins across a range appropriate for each observation epoch, then performed a separate four-component GMM analysis on each of these bins. Wd1 stars were identified as that GMM component which exhibited the smallest velocity dispersion, which was roughly an order of magnitude less than that of any of the field components. These per-magnitude maps were then completeness corrected using magnitude-wise completion maps derived from synthetic data as above. The corrected maps were recombined to reconstruct the desired field and Wd1 maps. Results were compared with a four-component GMM formed from all stars in the observation field, with correction again applied on a magnitude-by-magnitude basis. We found small and inconsistent differences across epochs and magnitudes for the structural similarity index of the field map with respect to a flat image between the results of these two GMM procedures. We interpret this as indicating that the GMM decomposition into field versus Wd1 stars of the VPD is consistent across magnitudes and that there is no benefit in performing magnitude-by-magnitude GMM modeling with this data set.

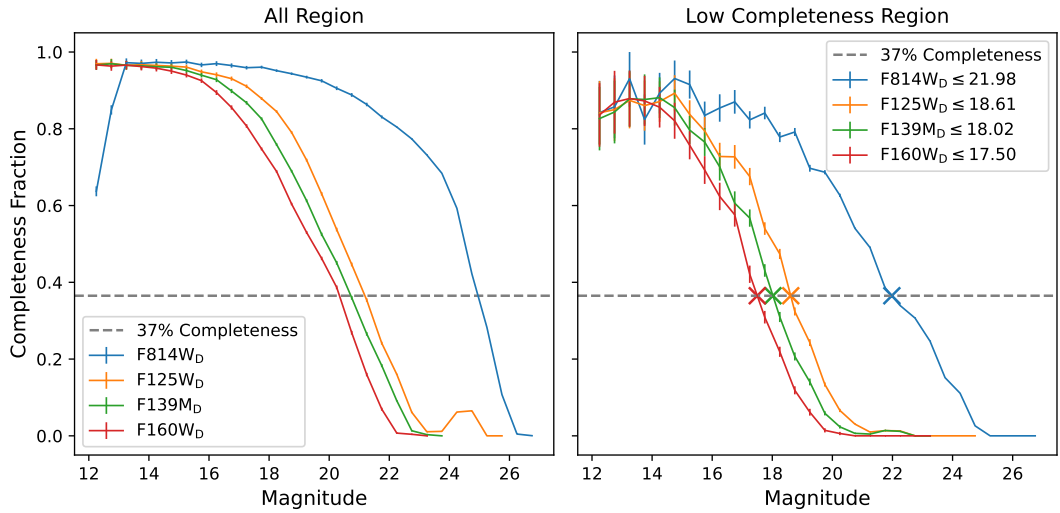

**Figure 11.** Completeness as a function of magnitude. Left: 1D completeness curves for the entire field. Right: 1D completeness curves for the low-completeness region marked in Figure 10. The gray dotted line in each plot marks the 65% completeness limit. After masking out the lowest 15% completeness regions, our cut at 17.5 mag in dereddened F160W$_D$ is equivalent to a 65% completeness limit in the low-completeness region.

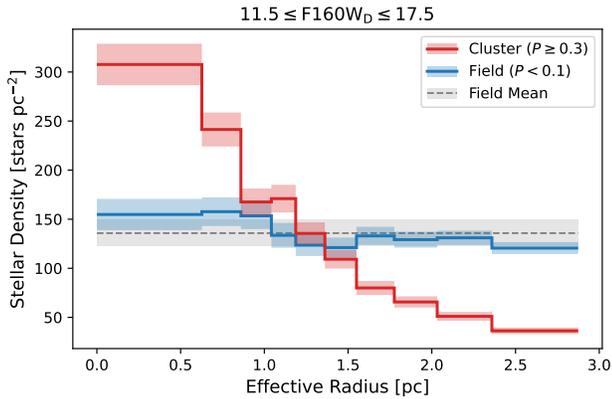

**Figure 12.** Dependence of completeness-corrected cluster and field stellar densities on effective radius. Completeness is interpolated as a function of effective radius and magnitude. The cluster members with $p_{\rm clust} \geqslant 0.3$ are shown in red, and the field stellar density with $p_{\rm clust} < 0.1$ is shown in blue. The gray dashed line and shaded region represent the mean field stellar density across the effective radius.

### 4.4. Stellar Density Modeling

Characterizing the stellar surface density and radial density profile of Wd1 requires accounting for both its known eccentricity (e.g., M. P. Muno et al. 2006; W. Brandner et al. 2008; M. Gennaro et al. 2011) and the edge effects introduced by the limited FOV.

To quantify the elongation and orientation of the cluster, we fit a 2D Gaussian profile to the stellar surface density map, using stars with $p_{\rm clust} \geqslant 0.3$ and F160W$_D$ between 11.5 and 17.5 mag. Membership and completeness correction are incorporated in the surface stellar density by weighting each star with

$$w_i = \frac{p_\mu^i}{C_i}, \quad (11)$$

with $p_\mu^i$ denoting the PM membership of the $i$th star, and $C_i$ its completeness. The stellar density map is computed on a pixel grid. Specifically, the density at each pixel is calculated as the sum of the weights of the 50 nearest stars from that pixel divided by the overlapping area between the masked image and a circle centered on the pixel, with its radius being the distance to the 50th nearest neighbor.

Based on the best-fit Gaussian profile parameters, we define the effective radius:

$$r_{\rm eff} = \sqrt{(\Delta x/\epsilon)^2 + \Delta y^2}, \quad (12)$$

where $\Delta x$ and $\Delta y$ are the offsets from the Gaussian center projected along the minor and major axes, respectively, and $\epsilon$ is the ratio of the minor-to-major axes, or ellipticity. We note that the definitions of ellipticity and eccentricity vary in the literature. In this work, we adopt the eccentricity defined as $e = \sqrt{1 - \epsilon^2}$. The effective radius transforms stellar positions by scaling them along the minor axis to match the major axis of the Gaussian profile, thereby circularizing the spatial distribution and enabling radial analysis.

The limited FOV affects the accurate determination of the radial density profile. To account for the edge effect, we introduce the concept of area fraction $f(r_{\rm eff})$ as a function of the effective radius. The area fraction adjusts for the incomplete radial coverage near the edges. Specifically, consider an elliptical annulus with an effective radius of $r_{\rm eff}$ and a width of $dr_{\rm eff}$ oriented concentrically with the Gaussian surface density profile. The area fraction is defined as the ratio of the remaining area of the annulus within the completeness-masked image to the full area of the annulus, assuming an infinite FOV:

$$f(r_{\rm eff}) = \frac{A_{\rm overlap}(r_{\rm eff})}{A_{\rm total}(r_{\rm eff})}, \quad (13)$$

where $A_{\rm overlap}(r_{\rm eff})$ is the remaining area and $A_{\rm total}(r_{\rm eff}) = 2\pi r_{\rm eff} dr_{\rm eff}$ is the total area of the annulus, respectively. Despite the areas being calculated in stretched coordinates with $r_{\rm eff}$, the ratio remains identical in the original coordinates. We interpolate the area fraction as a function of the effective radius for annuli with a width increment of 1 pixel and concentric with the Gaussian profile.

Each star is then weighted by $w_i^{r_{\rm eff}}$ to determine the radial density profile:

$$w_i^{r_{\rm eff}} = \frac{1}{f(r_{\rm eff})} \frac{p_\mu^i}{C_i}. \quad (14)$$

This weight is only used in radial profile modeling. To prevent amplifying the weights beyond reasonable levels, we cautiously restrict the area fraction $f(r_{\rm eff}) \geqslant 0.3$.

Next, we fit the radial profile with an Elson–Fall–Freeman (EFF) model (R. A. W. Elson et al. 1987), which is a good description of other young clusters (A. D. Mackey & G. F. Gilmore 2003a, 2003b; D. E. McLaughlin & R. P. van der Marel 2005; M. W. Hosek et al. 2015; N. Z. Rui et al. 2019) as well as Wd1 (M. Gennaro et al. 2011; M. Andersen et al. 2017). The EFF profile takes the form

$$\Sigma(r_{\rm eff}) = \Sigma_0 \left(1 + \frac{r_{\rm eff}^2}{a^2}\right)^{-\gamma/2}, \quad (15)$$

where $\Sigma_0$ is the amplitude, $r$ is the radius, $a$ is the core parameter, and $\gamma$ is the slope of the power law. We assume our background contamination is zero, as the radial range explored in this work is insufficient to constrain the parameter robustly. The background contamination parameter is not constrained if included, and affects the convergence of other parameters due to degeneracies in the parameter space. The core parameter $a$ is related to the core radius $r_c$ of the cluster by

$$r_c = \frac{a}{\sqrt{2^{2/\gamma} - 1}}. \quad (16)$$

Note that we only model $r_c$, and the posterior of $a$ is purely converted from $r_c$ and shown for illustration purposes.

To perform the EFF fit, we use MultiNest to maximize the log-likelihood function:

$$\log \mathcal{L} = \sum_{i=1}^{N} w_i^{r_{\rm eff}} \log \Sigma(r_{\rm eff}), \quad (17)$$

where $N$ is the number of stars used for the fit, $w_i^{r_{\rm eff}}$ is the weight of each star defined in Equation (14), and $\Sigma(r_{\rm eff})$ is the EFF model. We refer to J. C. Richardson et al. (2011) and L. Cicuéndez et al. (2018) for the derivation and further discussion of the log-likelihood function and T. Do et al. (2013) and M. W. Hosek et al. (2015) for further discussion of the methodology.

### 4.5. Color–Magnitude Diagram Fitting

We performed a Bayesian approach identical to M. W. Hosek et al. (2019) to model the cluster parameters in CMD space using simulated synthetic clusters based on theoretical isochrones and infer the distance $d$ and age $t$ of the cluster. Here, we briefly summarize the modeling process and direct the readers to M. W. Hosek et al. (2019) for further details. We used SPISEA (M. W. Hosek et al. 2020) to generate the simulated model cluster with combined stellar evolutionary models, atmosphere models, a two-segment broken power-law IMF, and an initial–final mass relation. The IMF has a slope of $\alpha_1$ for stellar masses below the break-point mass $m_{\rm break}$, and a slope of $\alpha_2$ for masses above $m_{\rm break}$. The free parameters include the cluster distance $d$, age $t$, average extinction value $A_{Ks}$, differential extinction $dA_{Ks}$, IMF break-point mass $m_{\rm break}$, higher-mass IMF slope $\alpha 1$, the lower- to higher-mass IMF slope ratio $d\alpha = \alpha_2/\alpha_1$, break mass $m_{\rm break}$, and the cluster mass $M_{\rm cl}$. Specifically, stars were sampled from the IMF accounting for observationally unresolved multiplicity according to the parameters outlined in J. R. Lu et al. (2013), with a total mass of $M_{\rm sim} = 5 \times 10^6 \, M_\odot$ to reduce stochastic variations. The stellar physical properties were determined using the merged set of stellar evolutionary models in SPISEA as described in Section 4.2. The derived stellar parameters were used as input for the same suite of stellar atmospheric models mentioned in Section 4.2 to generate the synthetic photometry of the stars using pysynphot (STScI Development Team 2013). Each star in the simulated cluster was assigned a radius following the same distribution as the observed effective radius $r_{\rm eff}$ in Wd1.

The posterior probability function of the cluster parameters is given by the Bayesian theorem:

$$\mathcal{L}(\theta) = p(\boldsymbol{k}, N|\theta) \cdot p(\theta), \quad (18)$$

where $\boldsymbol{k} = \{(m_i, c_{1,i}, c_{2,i})\}_{i=1}^{N}$ denotes the CMD positions of $N$ observed stars in F160W$_D$ magnitudes, F814W$_D$ − F125W$_D$ color, and F125W$_D$ − F160W$_D$ color, respectively. $p(\theta)$ is the prior function for the model parameters. The model parameter vector $\theta$ includes the cluster distance $d$, age $t$, extinction $A_{Ks}$, differential extinction $\Delta A_{Ks}$, total cluster mass $M_{\rm cl}$, as well as the break mass and slopes of the IMF. The likelihood function $p(\boldsymbol{k}, N|\theta)$ consists of two components:

$$p(\boldsymbol{k}, N|\theta) = p(\boldsymbol{k}|\theta) \cdot p(N|\theta), \quad (19)$$

where $p(\boldsymbol{k}|\theta)$ is the probability of the observed CMD given the model, and $p(N|\theta)$ is the Poisson probability of observing $N$ stars given the expected number of stars $N_{\rm exp}$ of the model. The CMD likelihood is calculated as the product of the CMD probability of each star:

$$p(\mathbf{k}|\theta) = \prod_{i=1}^{n} p(\mathbf{k}_i|\theta). \quad (20)$$

For each star, we modeled its CMD probability as a multivariate normal distribution centered at the observed values with the standard deviation being the uncertainties $\mathcal{N}(\mathbf{k}_i, \boldsymbol{\sigma}_i)$. The individual stellar CMD likelihood is then calculated as the normal distribution convolved with a mixture of cluster and field model cluster CMD probability density functions (PDFs) weighted by the corresponding membership probability:

$$p(\mathbf{k}_i|\theta) = \int \mathcal{N}(\mathbf{k}; \mathbf{k}_i, \boldsymbol{\sigma}_i)[p_\mu \cdot f_{\rm cl}(\mathbf{k}_{\rm sim}|\theta) \\ + (1 - p_\mu) \cdot f_{\rm field}(\mathbf{k}_{\rm field})] \, d\mathbf{k}, \quad (21)$$

where $\mathcal{N}(\mathbf{k}; \mathbf{k}_i, \boldsymbol{\sigma}_i)$ denotes the PDF of the multivariate normal distribution with a mean of $\mathbf{k}_i$ and a standard deviation of $\boldsymbol{\sigma}_i$, $p_\mu$ is the PM membership probability, $f_{\rm cl}(\mathbf{k}_{\rm sim}|\theta)$ is the CMD PDF weighted by the completeness as a function of effective radius $C(r_{\rm eff})$, and $f_{\rm field}$ is the CMD PDF of the field contamination population.

The total number likelihood is given by

$$p(N|\theta) = \frac{(N_{\rm exp})^N \exp(-N_{\rm exp})}{N!}, \quad (22)$$

where $N$ is the number of observed stars, and $N_{\rm exp}$ is the number of expected stars given $\theta$. The expected number of stars is determined by linearly scaling the completeness-corrected number of stars in the simulated cluster by the ratio of the observed cluster mass to the simulated cluster mass:

$$N_{\rm exp} = N \times \frac{M_{\rm cl}}{M_{\rm sim}}. \quad (23)$$

With the likelihood function, we used emcee to sample the posterior distribution of the model parameters with 100 walkers, 300 steps, and KDEMove. The results are presented in Section 5.3.

## 5. Results

We present the full cluster catalog with PMs, dereddened magnitudes, kinematic and color membership probabilities, and completeness values of each star in Table 4. In this section, we use the catalog to model and analyze the properties of Wd1, including its surface density, morphology, radial density profile, age, distance, velocity dispersion, and mass segregation.

### 5.1. Surface Density and Morphology

We calculate the surface stellar density map of stars with $p_{\rm clust} \geqslant 0.3$ and F160W$_{\rm D}$ between 11.5 and 17.5 mag with membership and completeness correction. This results in a star count of $N = 1951$.

To determine the morphology of Wd1, including its orientation, centroid, and elongation, we fit a 2D Gaussian profile to the membership and completeness-corrected stellar density map as described in Section 4.4. Throughout this work, we adopted the center of the Gaussian density profile as the Wd1 center, located at R.A. = $16^{\rm h}47^{\rm m}04\overset{\rm s}{.}0$, decl.=$-45°51'04\overset{''}{.}7$ (J2000). We find that the cluster is elongated in the northeast–southwest direction, with the major axis at a position angle of $\sim 35°.2$ east of north. This aligns the elongation with the Galactic plane and cluster PM movement spread, as seen in Figure 5. The flattening or ellipticity of the Gaussian profile $\epsilon$, defined as the ratio of the minor to the major axis, is approximately 0.70, translating to an eccentricity of $e = \sqrt{1 - \epsilon^2} = 0.71$.

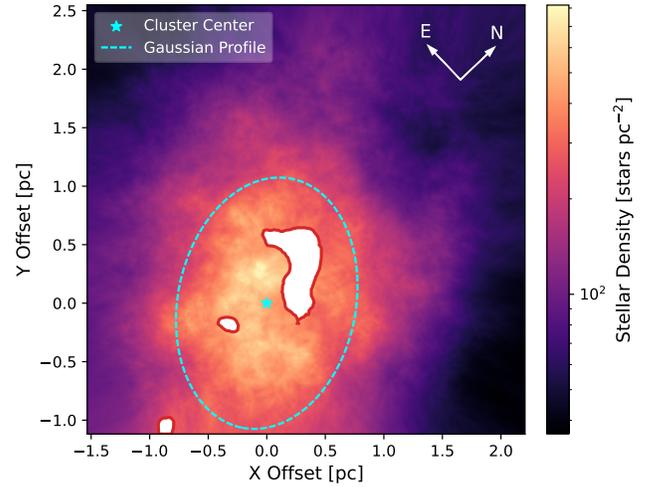

**Figure 13.** Stellar density map corrected by membership probability and completeness for stars with $p_{\rm clust} \geqslant 0.3$ and F160W$_{\rm D}$ between 11.5 and 17.5 mag. The cyan star and dashed ellipse indicate the cluster center and the best-fit $1\sigma$ Gaussian profile, respectively. Note that the centroid has shifted to the lower left, or south. This is due to the large extinction values in that area being accounted for in this map.

The stellar density map and the Gaussian profile of the full catalog and three mass bins are shown in Figures 13 and 14, respectively. The cyan star represents the cluster center, and the dashed ellipse marks the $1\sigma$ contour of the 2D Gaussian density profile. Both figures adopt a log-scale color map. The elongation and its direction can be observed from Figure 13. The decrease in size with increasing mass is clearly visible in Figure 14 under an identical color map for each mass bin, indicating that the massive stars are more concentrated in the center. However, we note that the determination of the cluster morphology is subject to the limited FOV and the masked low-completeness region. A broader FOV or more complete observation is required to rigorously characterize the cluster morphology.

### 5.2. Radial Density Profile

We derive the best-fit models based on the entire catalog and divide the full sample into three mass bins with a total weight ratio of roughly 4:3:2 from the lowest to highest mass to explore the difference in radial density for each mass bin. The number of observed and weighted stars under each criterion and each mass bin is summarized in Table 5.

Together with the completeness magnitude cut at F160$_{\rm D} \leqslant 17.50$, or equivalently $M \geqslant 1.12 \, M_\odot$, under the best-fit isochrone to be discussed in Section 5.3, we ended up with a maximum effective radius of 2.87 pc. The resulting catalog has 1887 uncorrected stars, or 2304.6 stars after being weighted by Equation (14).

Figure 15 shows the measured stellar density radial profiles and their corresponding best-fit EFF models. Radial profiles are measured in 15 radial bins that share the same center, orientation, and elongation with the Gaussian density profile,

Table 4
Photometric, Kinematic, and Structural Properties of Westerlund 1

| ID | F814W$_D$ (mag) | F125W$_D$ (mag) | F160W$_D$ (mag) | $\Delta\alpha_0^{*a}$ (arcsec) | $\Delta\delta_0$ (arcsec) | $\mu_{\alpha^*}$ (mas yr$^{-1}$) | $\mu_\delta$ (mas yr$^{-1}$) | $t_0$ (yr) | $p_\mu$ | $p_{color}$ | $p_{clust}$ | C | M ($M_\odot$) | $r_{eff}$ (pc) | $f(r_{eff})$ |
|---|---|---|---|---|---|---|---|---|---|---|---|---|---|---|---|
| wd1 00001 | 18.50 | 15.17 | 14.29 | −29.13 | −17.96 | −0.49 ± 0.14 | −0.19 ± 0.15 | 2011.698 | 0.30 | 1 | 0.30 | 0.97 ± 0.21 | 4.86 | 0.67 | 0.87 |
| wd1 00002 | 18.88 | 15.41 | 14.50 | −53.11 | 66.75 | 0.53 ± 0.04 | 0.35 ± 0.04 | 2010.997 | 0.59 | 1 | 0.59 | 0.97 ± 0.18 | 4.40 | 2.15 | 0.61 |
| wd1 00003 | 17.54 | 14.57 | 13.83 | 3.09 | 149.75 | −0.01 ± 0.05 | 0.21 ± 0.04 | 2009.482 | 0.89 | 1 | 0.89 | 1.00 ± 0.19 | 5.99 | 3.11 | 0.18 |
| wd1 00004 | 18.51 | 15.28 | 14.45 | 98.79 | 26.37 | 0.06 ± 0.02 | 0.10 ± 0.02 | 2008.278 | 0.92 | 1 | 0.92 | 0.99 ± 0.18 | 4.51 | 2.20 | 0.58 |
| wd1 00005 | 19.50 | 15.80 | 14.84 | 40.85 | 61.87 | 0.10 ± 0.04 | 0.25 ± 0.04 | 2009.879 | 0.90 | 1 | 0.90 | 0.96 ± 0.16 | 3.74 | 1.34 | 0.82 |
| ... | ... | ... | ... | ... | ... | ... | ... | ... | ... | ... | ... | ... | ... | ... | ... |

**Notes.** Description of columns: ID: star name. F814W$_D$, F125W$_D$, and F160W$_D$: magnitudes in the corresponding filters (Vega). $\Delta\alpha_0^*$, $\Delta\delta_0$: relative position in R.A. and decl. at time $t_0$. $t_0$: average observation time weighted by the inverse of the astrometric uncertainties. $p_\mu$: PM membership. $p_{color}$: color membership. $p_{clust}$: product of PM membership and color membership. C: completeness. M: stellar mass. $r_{eff}$: effective radius. $f(r_{eff})$: area fraction.

[a] Positions are relative to R.A. = 16$^h$47$^m$04$^s$.0, decl.=−45°51′04″.7 (J2000).

(This table is available in its entirety in machine-readable form in the online article.)

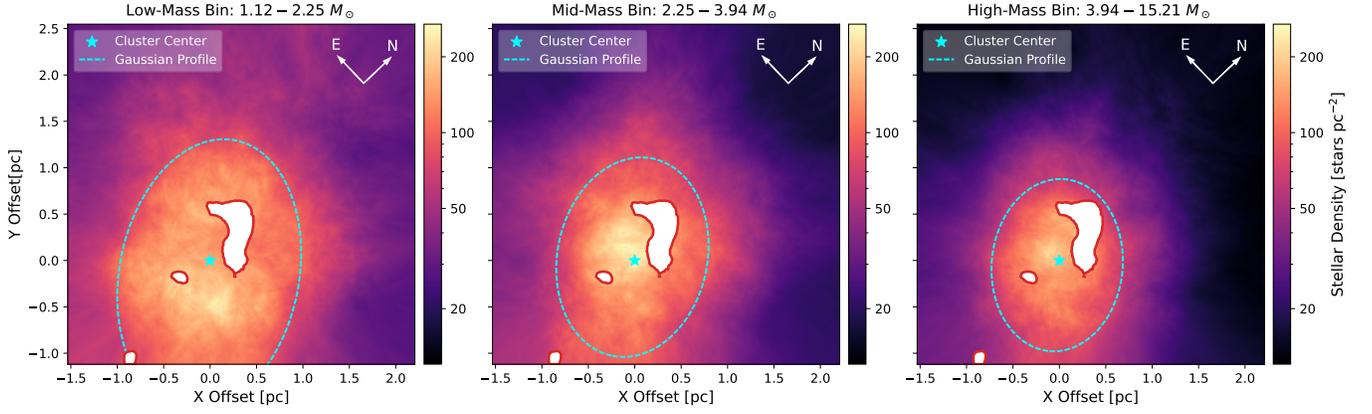

**Figure 14.** Stellar density map for increasing mass bins. The cyan star marks the center of the full cluster, and the dashed ellipse shows the $1\sigma$ Gaussian profile for each mass bin. Left: low-mass bin, $1.12 - 2.25\ M_\odot$. Middle: mid-mass bin, $2.25 - 3.94\ M_\odot$. Right: high-mass bin, $3.94 - 15.21\ M_\odot$. Note that the color map used in each panel is in log scale, which are consistent across panels.

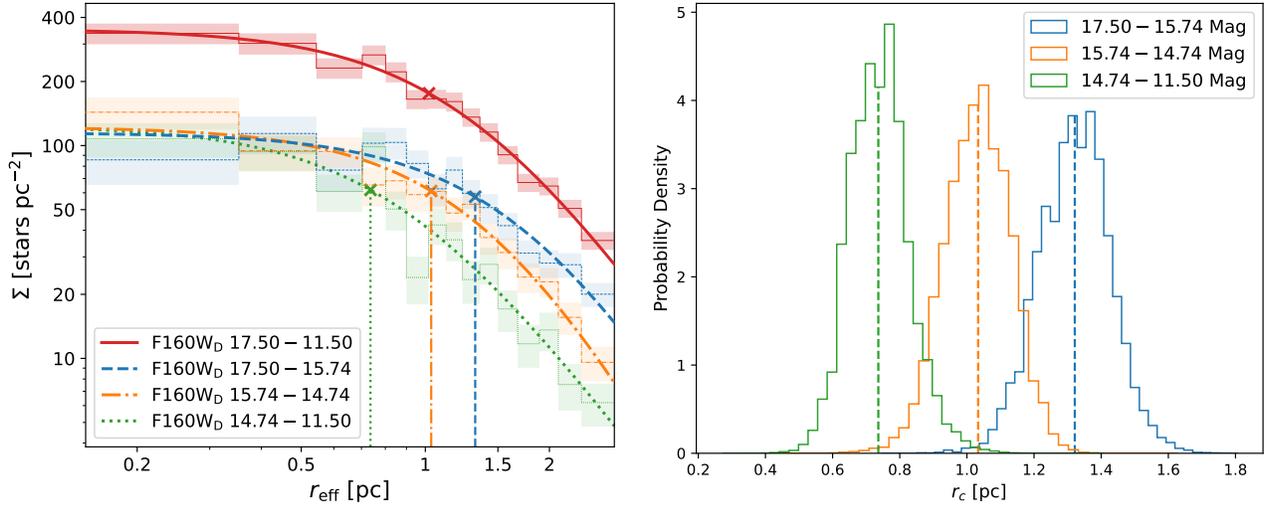

**Figure 15.** Left: binned radial profile and EFF model in each mass bin. The binned radial profiles are represented as stair plots, and the EFF models are shown as the curves. The width of each stair corresponds to the bin $r_\mathrm{eff}$ range, and the shaded region represents the Poisson uncertainty. Bins were chosen to have an equal number of stars weighted by $w_i^r$ with area fraction $f \geqslant 0.30$, which corresponds to $r_\mathrm{eff} \leqslant 2.87$ pc. Right: posterior distributions of $r_c$ with the median value marked with a dashed line.

and contain an equal number of weighted sources. It is worth mentioning that the observed values presented in the stair plot in Figure 15 are calculated as the sum of weights defined in Equation (14) divided by the actual area of each bin, which allows a more accurate representation of density in each bin. The area-fraction-corrected weight in Equation (14) is only used in EFF profile modeling. The posteriors of the EFF radial profile fits are shown in Figure 16 for the full and three binned catalogs. The aforementioned low-completeness region in Figure 10 is the innermost radial bin.

The results of the EFF profile model parameters are summarized in Table 6, including a comparison with the Arches and Quintuplet clusters, which will be discussed in Section 6.3. The core radius $r_c$ of the full cluster is $0.10^{+0.07}_{-0.06}$ pc, and it decreases with increasing mass, indicating mass segregation.

### 5.3. Age and Distance

With the posterior probability function described in Section 5.3, the best-fit cluster properties are determined utilizing the Markov Chain Monte Carlo ensemble sampler emcee (D. Foreman-Mackey et al. 2013). We utilized 100 samplers, 300 steps, and the KDEMove proposal in emcee. The final 100 converged steps are considered as the burn-in steps and are used to determine the value and uncertainties of the parameters. We modeled the population with high-membership probability $p_\mathrm{clust} \geqslant 0.3$ and within the completeness magnitude range of $17.5 \leqslant F160W_D \leqslant 11.5$. In this work, we focus on the constraints from the cluster age and distance, which are relevant for later discussions on velocity dispersion (Section 5.4) and mass segregation (Section 5.5). The related free parameters include the distance $d$, age $t$, average extinction $A_{Ks}$, and differential extinction $dA_{Ks}$. Their prior ranges and modeling results are summarized in Table 7. Additional free parameters in CMD modeling, such as the IMF slopes, IMF peak mass, and cluster mass, will be constrained more robustly with new observational data in an upcoming paper (L. Wei et al. 2025, in preparation). We binned the CMD of the observed and best-fit simulated clusters and their difference for visualization comparisons in Figure 17. The marginalized posteriors are shown in Figure 18. The resulting best-fit cluster distance is $3723.77 \pm 113.28$ pc, and the age is $\log_{10} t = 6.87 \pm 0.03$, corresponding to $7.45 \pm 0.53$ Myr. Notice that there is a degeneracy in the distance–age posterior, as shown in Figure 18. The distance and age have a negative correlation, with $d$ approximately spanning between 3.5 and 4 kpc, and $\log_{10} t$ ranging roughly from 7 to 6.6. The reported uncertainties correspond to half of the difference between the 16th and 84th percentiles of the burn-in samplers. The best-fit cluster age and distance were used to generate the isochrone shown in Figure 9 and interpolate the stellar masses. Note that the stellar mass estimate might be slightly affected by the solar metallicity assumption.

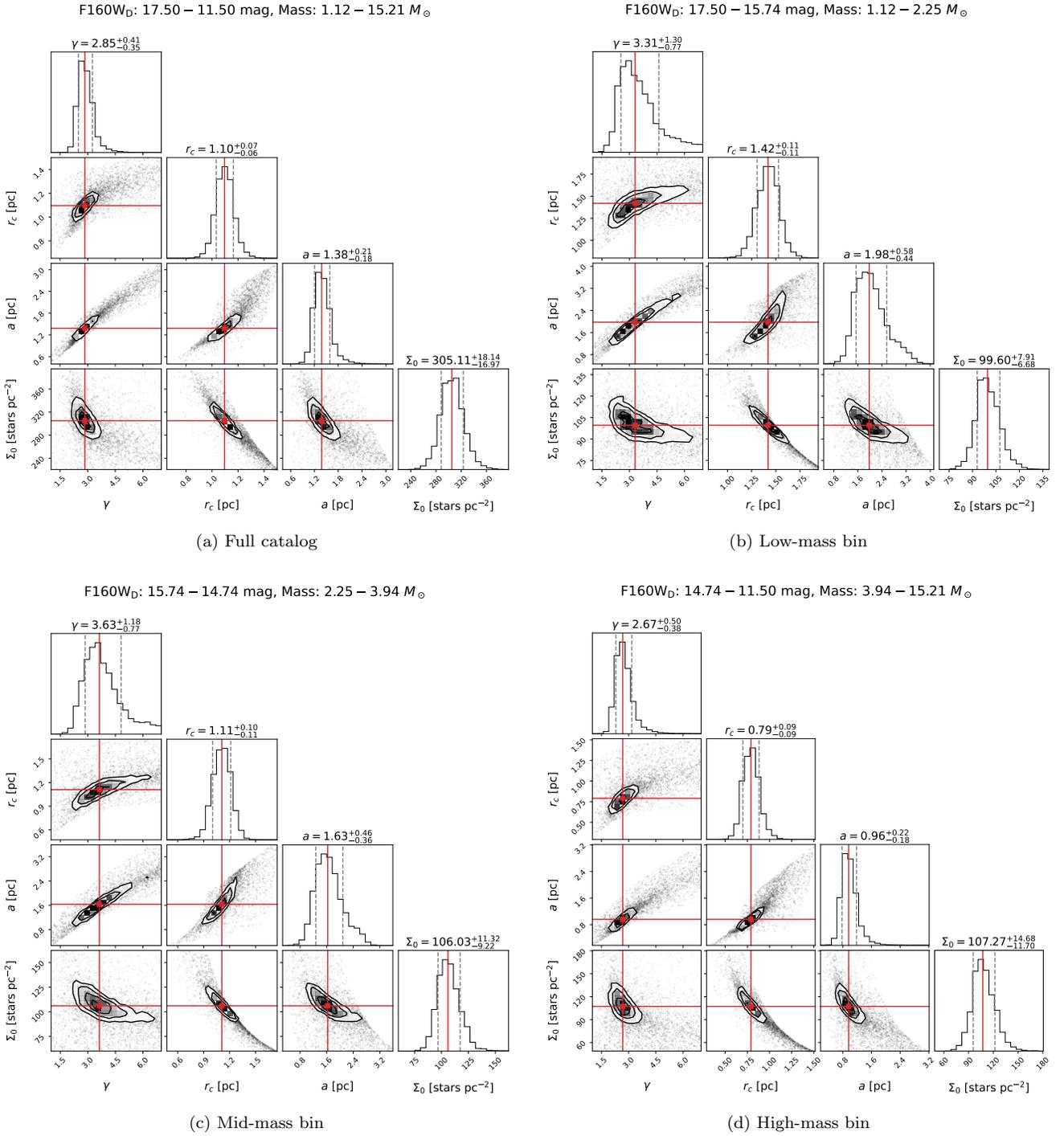

**Figure 16.** Weighted posterior distributions of EFF radial profile parameters. (a) Full catalog. (b) Low-mass bin. (c) Mid-mass bin. (d) High-mass bin. The red lines mark the weighted median, and the gray dashed line marks the weighted 16th and 84th percentiles. Note that we only modeled $r_c$, and the posterior distribution of $a$ is purely converted from $r_c$.

Although the discrepancy between the PMS star colors and the isochrone model mentioned in Section 4.2 is not included in the modeling because of the completeness limit of $F160W_D \leqslant 17.5$, we tried extending the magnitude limit down to 18.5 to see if it helps mitigate the color difference. The result is negative, and the sampler still converges close to our reported values. Upon inspection, we find that none of the model isochrones display such a blue color at the PMS stage, regardless of distance or age. We will leave exploration of different stellar evolutionary and atmospheric models for a future paper. However, we emphasize again that this caveat does not affect the structural and kinematics analysis in this work, as the problematic magnitude range is excluded from the modeling due to our completeness limit.

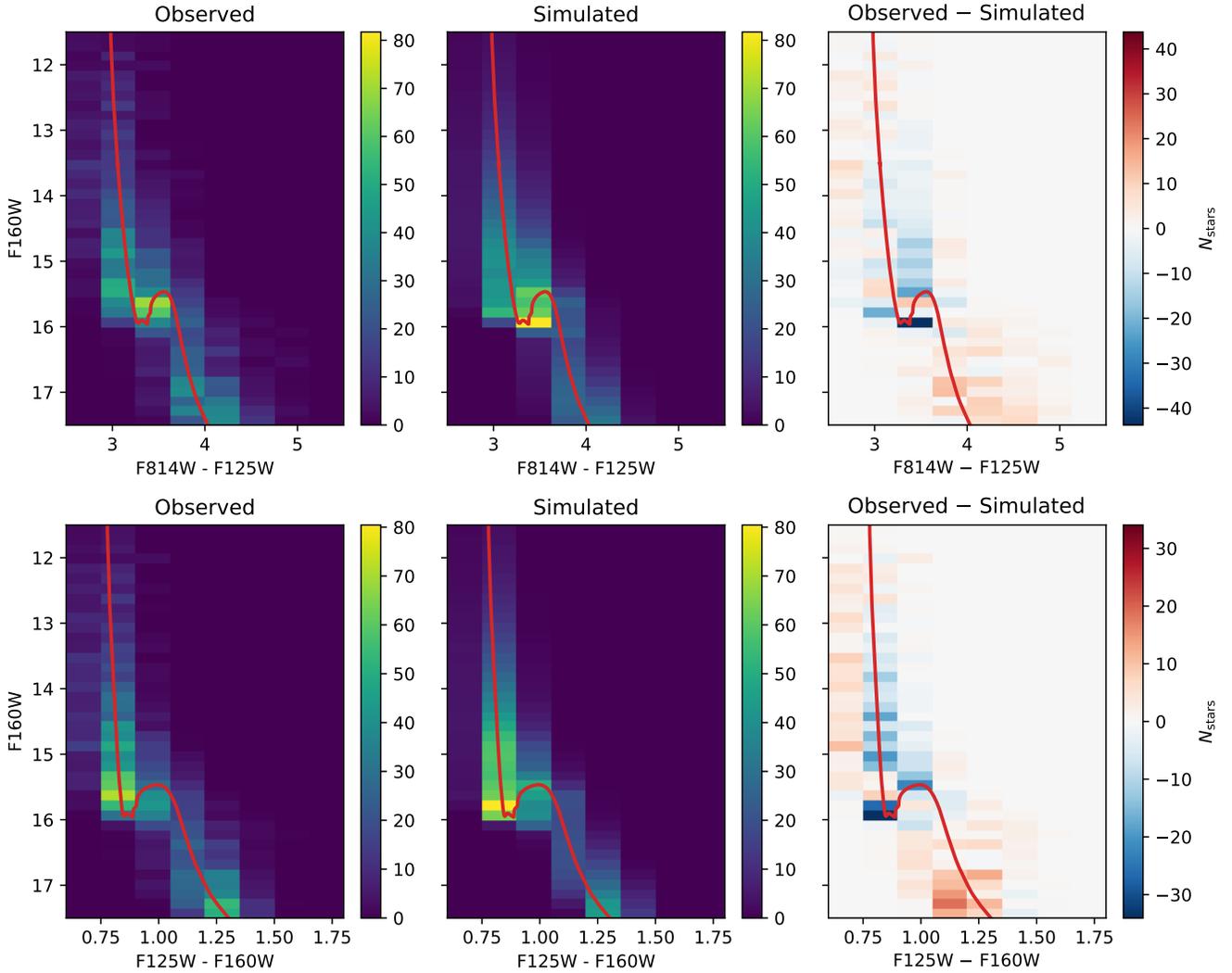

**Figure 17.** Observed and best-fit simulated cluster binned CMD and the residuals. The top row shows the $F814W_D - F125W_D$ color and the bottom row shows the $F125W_D - F160W_D$ color. The left, middle, and right columns correspond to the observed CMD, simulated CMD, and the residuals, respectively. The best-fit isochrone is overplotted as the red curve. The left and middle columns share the same color map range in each row.

### 5.4. Velocity Dispersion

We model the intrinsic PM velocity dispersion of the subsample by a mixture of two bivariate Gaussians, one for the cluster members (subscripted by "cl") and the other for the residual foreground stars (subscripted by "fg"). Each Gaussian is characterized by the mean PM values $\bar{\boldsymbol{\mu}} \equiv (\bar{\mu}_R, \bar{\mu}_T)$, the PM dispersions $\boldsymbol{\sigma} \equiv (\sigma_R, \sigma_T)$, and the correlation coefficient $\rho$. Given a set of PM measurements $D \equiv \{\boldsymbol{\mu}_i, \boldsymbol{\epsilon}_i\}_{i=1}^N$, where $\boldsymbol{\mu}_i \equiv (\mu_{R,i}, \mu_{T,i})$ is the PM of the $i$th star and $\boldsymbol{\epsilon}_i \equiv (\epsilon_{R,i}, \epsilon_{T,i})$ are the associated uncertainties, the posterior for the Gaussian parameters $\boldsymbol{p}_{\rm cl} \equiv (\bar{\boldsymbol{\mu}}_{\rm cl}, \boldsymbol{\sigma}_{\rm cl}, \rho_{\rm cl})$ and $\boldsymbol{p}_{\rm fg} \equiv (\bar{\boldsymbol{\mu}}_{\rm fg}, \boldsymbol{\sigma}_{\rm fg}, \rho_{\rm fg})$ is:

$$P(\boldsymbol{p}_{\rm cl}, \boldsymbol{p}_{\rm fg} | D) \propto P(D | \boldsymbol{p}_{\rm cl}, \boldsymbol{p}_{\rm fg}) P(\boldsymbol{p}_{\rm cl}, \boldsymbol{p}_{\rm fg}), \quad (24)$$

according to Bayes' theorem, where $P(D|\boldsymbol{p}_{\rm cl}, \boldsymbol{p}_{\rm fg})$ is the likelihood and $P(\boldsymbol{p}_{\rm cl}, \boldsymbol{p}_{\rm fg})$ is the prior. The likelihood $P(D|\boldsymbol{p}_{\rm cl}, \boldsymbol{p}_{\rm fg})$ is a product of the two Gaussian mixtures for each data point:

$$P(D|\boldsymbol{p}_{\rm cl}, \boldsymbol{p}_{\rm fg}) \propto \prod_i [(1-\eta)\,\mathcal{N}(\boldsymbol{\mu}_i; \bar{\boldsymbol{\mu}}_{\rm cl}, \Sigma_{{\rm cl},i}) + \eta\,\mathcal{N}(\boldsymbol{\mu}_i; \bar{\boldsymbol{\mu}}_{\rm fg}, \Sigma_{{\rm fg},i})], \quad (25)$$

where $\eta$ is the fraction of foreground contaminants, and $\mathcal{N}(\boldsymbol{\mu}_i; \bar{\boldsymbol{\mu}}, \Sigma_i)$ denotes the PDF of a bivariate normal distribution characterized by a mean of $\bar{\boldsymbol{\mu}}$ and covariance of $\Sigma_i$ evaluated for the $i$th star with measured PMs $\boldsymbol{\mu}_i$. The covariance matrices of the $i$th star $\Sigma_{{\rm cl},i}$ and $\Sigma_{{\rm fg},i}$ are associated with the measurement errors and intrinsic kinematics of the cluster and the foreground population, respectively:

$$\Sigma_{{\rm cl},i} = \begin{bmatrix} \sigma_{R,{\rm cl}}^2 + \epsilon_{R,i}^2 & \rho_{\rm cl}\,\sigma_{R,{\rm cl}}\,\sigma_{T,{\rm cl}} \\ \rho_{\rm cl}\,\sigma_{R,{\rm cl}}\,\sigma_{T,{\rm cl}} & \sigma_{T,{\rm cl}}^2 + \epsilon_{T,i}^2 \end{bmatrix}, \quad (26)$$

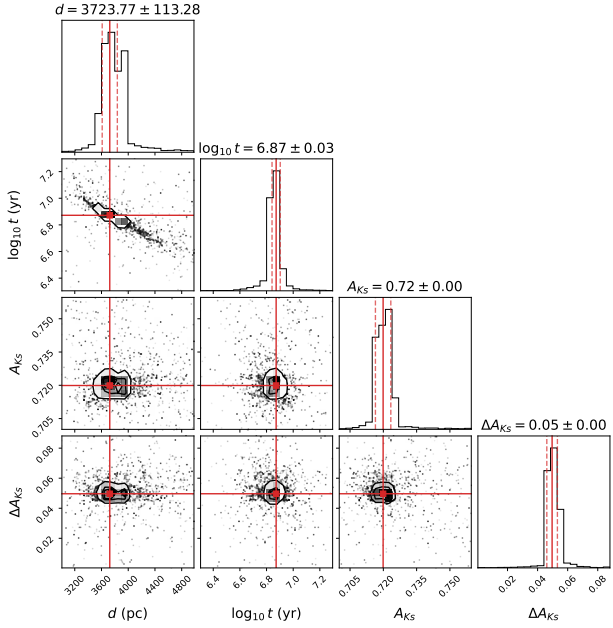

**Figure 18.** Posteriors of the CMD fit. The red solid lines mark the median of the burn-in samplers, and the red dashed lines represent the 16th and 84th percentiles, or 1σ uncertainties.

$$\Sigma_{\text{fg},i} = \begin{bmatrix} \sigma_{R,\text{fg}}^2 + \epsilon_{R,i}^2 & \rho_{\text{fg}}\, \sigma_{R,\text{fg}}\, \sigma_{T,\text{fg}} \\ \rho_{\text{fg}}\, \sigma_{R,\text{fg}}\, \sigma_{T,\text{fg}} & \sigma_{T,\text{fg}}^2 + \epsilon_{T,i}^2 \end{bmatrix}. \quad (27)$$

In the case of the prior $P(\boldsymbol{p}_{\text{cl}}, \boldsymbol{p}_{\text{fg}})$, we adopt the noninformative Jeffreys priors for each bivariate Gaussian as follows (J. M. Bernardo & F. J. Girón 1988; J. O. Berger & D. Sun 2008):

$$P(\boldsymbol{p}_{\text{cl}}, \boldsymbol{p}_{\text{fg}}) \propto \frac{1}{\sqrt{(1-\eta)\,\eta}} \prod_{k=\text{cl,fg}} [\sigma_{R,k}\, \sigma_{T,k}(1-\rho_k^2)]^{-2}. \quad (28)$$

We restrict the velocity dispersion modeling to a subsample of member candidates with high-quality PM measurements. The selection criteria are listed as follows:

i. synthetic PM measurement errors smaller than 0.2 mas yr$^{-1}$;
ii. color membership cuts illustrated Figure 6; and
iii. dereddened F160W$_D$ brighter than 16 mag.

The kinematic membership probabilities are not involved in the subsampling process, which would otherwise bias the velocity dispersions in favor of smaller values. Instead, criteria (ii) and (iii) are adopted to minimize the fraction of foreground stars while ensuring completeness across the FOV. By these criteria, 1204 stars are selected, whose PMs are then transformed from $x$–$y$ coordinates to radial ($R$) and tangential ($T$) components.

We sample the posterior $P(\boldsymbol{p}_{\text{cl}}, \boldsymbol{p}_{\text{fg}}|D)$ with emcee (D. Foreman-Mackey et al. 2013) and present the marginalized distributions of each parameter in Figure 19. Note that using uniform priors instead of the Jeffreys priors does not significantly impact our result. The fraction of foreground contaminants is estimated as low as $\sim 10\% \pm 2\%$.

The mean PM in radial and tangential components is $(\mu_{R,\text{cl}}, \mu_{T,\text{cl}}) = (0.03 \pm 0.01, -0.01 \pm 0.01)$ mas yr$^{-1}$, translating into $(0.57 \pm 0.22, -0.12 \pm 0.21)$ km s$^{-1}$ at a distance of 3723.77 pc, with positive tangential component corresponding to the counterclockwise direction. The radial component is consistent with zero within 3σ, indicating the positive value is statistically insignificant. The radial PM is not found to increase with radius from the cluster center either. Therefore, we find no evidence of expansion or contraction in the cluster, favoring the static scenario proposed by M. Gennaro et al. (2017).

The velocity dispersion measures $(\sigma_{R,\text{cl}}, \sigma_{T,\text{cl}}) = (0.25 \pm 0.01, 0.22 \pm 0.01)$ mas yr$^{-1}$, equivalent to $(4.33 \pm 0.19, 3.91 \pm 0.16)$ km s$^{-1}$ at the best-fit heliocentric distance 3723.77 pc. The 1D velocity dispersion is

$$\sigma_{\text{1D}} = \sqrt{\frac{\sigma_{R,\text{cl}}^2 + \sigma_{T,\text{cl}}^2}{2}} = 4.13 \pm 0.13 \text{ km s}^{-1}, \quad (29)$$

or equivalently $0.23 \pm 0.01$ mas yr$^{-1}$. We compare the velocity dispersion with the virial equilibrium model in Section 6.2.

### 5.5. Mass Segregation

We investigate mass segregation in the cluster with two metrics: the mass segregation ratio and the core radii of the radial profiles from Section 5.2.

We model mass segregation in the cluster using the mass segregation ratio, $\Lambda_{\text{MSR}}$, developed by R. J. Allison et al. (2009), which is recognized as a reliable approach with the advantage of not requiring a center or overall stellar distribution (see R. J. Parker & S. P. Goodwin 2015, for further discussion). This method examines the degree of mass segregation by the ratio of the minimum spanning tree (MST) length connecting the $N_{\text{MST}}$ most massive stars, $l_{\text{Massive}}$, to the mean MST length of $N_{\text{MST}}$ random stars, $\langle l_{\text{norm}} \rangle$. The distribution of $\langle l_{\text{norm}} \rangle$ is roughly Gaussian, and the error $\sigma_{\text{norm}}$ is then taken to be the standard deviation of the distribution. The mass segregation ratio $\Lambda_{\text{MSR}}$ is defined as

$$\Lambda_{\text{MSR}} = \frac{\langle l_{\text{norm}} \rangle}{l_{\text{Massive}}} \pm \frac{\sigma_{\text{norm}}}{l_{\text{Massive}}}. \quad (30)$$

When $\Lambda_{\text{MSR}}$ significantly differs significantly from one, there is either mass segregation or inverse mass segregation for values $> 1$ or $< 1$, respectively. A value of one indicates no mass segregation in the cluster.

We modify this method by finding the distribution of $l_{\text{Massive}}$ in a set of massive stars, rather than setting a fixed value of $N_{\text{MST}}$ massive stars. We randomly sample this set of massive stars in the same way we sample the random set of stars, and then compare the distribution of MST lengths. In this way, we can characterize the distribution of minimum separations of the massive stars more comprehensively. Our updated mass segregation ratio is defined as

$$\Lambda_{\text{MSR}} = \frac{\langle l_{\text{norm}} \rangle}{\langle l_{\text{Massive}} \rangle} \pm \sigma_{\text{MSR}}, \quad (31)$$

where

$$\sigma_{\text{MSR}} = \frac{\langle l_{\text{norm}} \rangle}{\langle l_{\text{Massive}} \rangle} \sqrt{\left(\frac{\sigma_{\text{norm}}}{\langle l_{\text{norm}} \rangle}\right)^2 + \left(\frac{\sigma_{\text{Massive}}}{\langle l_{\text{Massive}} \rangle}\right)^2}. \quad (32)$$

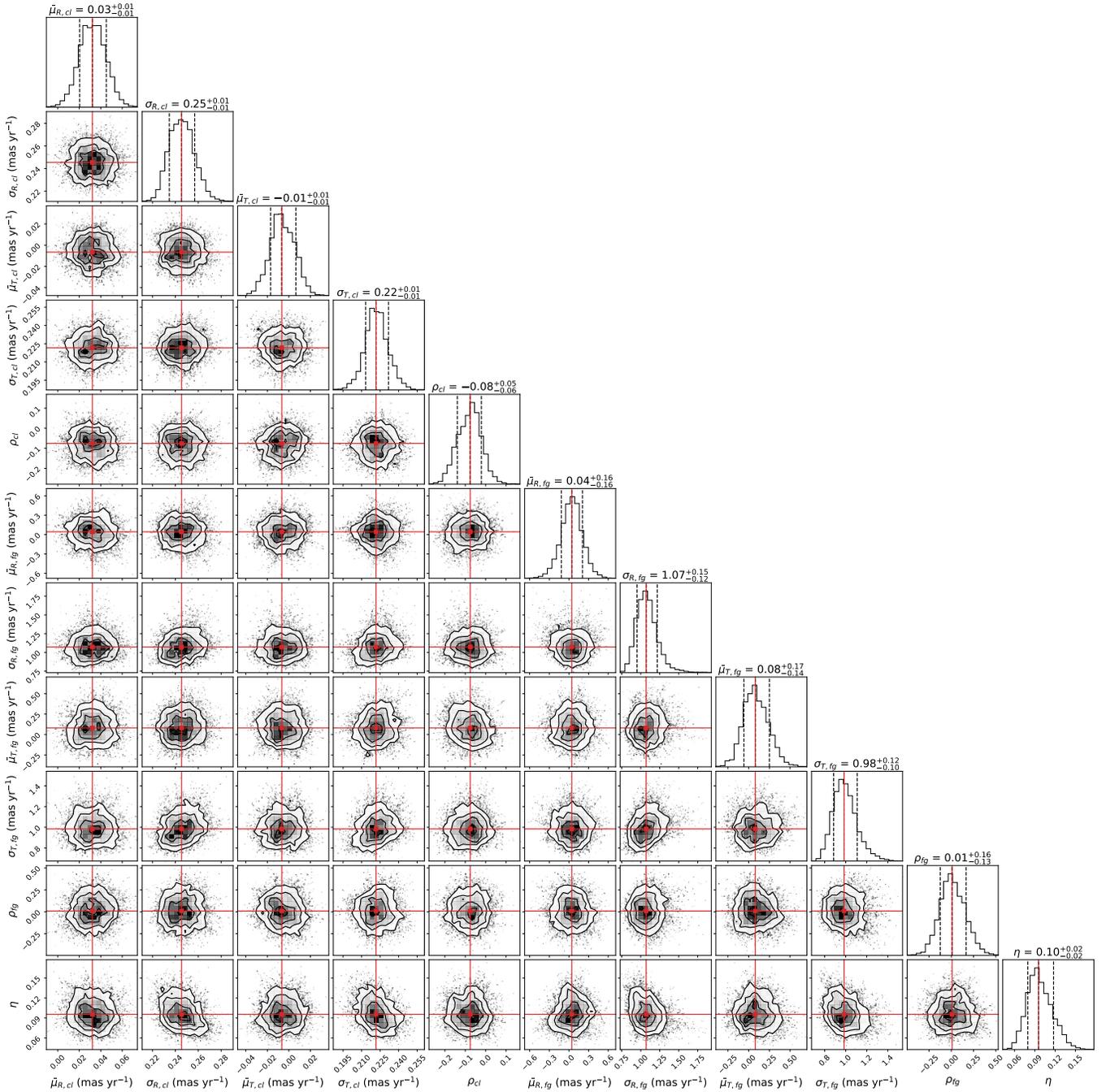

**Figure 19.** Marginalized posterior distributions for the parameters of kinematic modeling. The red line marks the median, and the dashed vertical lines mark the 16th and 84th percentiles.

We utilize `scipy` to construct the MST. To ensure a smooth distribution of MST lengths, we perform 5000 trials of 25 randomly chosen stars for both sets of stars. All populations are comprised of cluster members with $p_{\rm clust} \geq 0.3$. We account for the weight of each star by assigning the likelihood of selecting a given star proportional to its weight, as defined in Equation (11).

Using the 200 brightest stars with $F160W_D \leqslant 13.6$ ($\gtrsim 5.4\,M_\odot$) in our sample as the set of massive stars, we find $\Lambda_{\rm MSR} = 1.11 \pm 0.11$, slightly greater than one by only $1\sigma$. We also notice that the mean MST lengths tend to decrease as the mean stellar mass increases, as shown in Figure 20, where we show the distribution of MST lengths normalized by the mean of the full catalog for the three mass bins. Figure 21 illustrates the MST of each mass bin closest to the median of their corresponding tree lengths.

A similar trend is found in the posterior distributions of $r_c$ in the EFF radial profile modeling, as shown in Figure 15. Two-sided K-S tests comparing the radii of the stars in each mass bin also indicate that they are drawn from different parent distributions. The $p$-values corresponding to the low- to intermediate-, low- to high-, and intermediate- to high-mass bin comparisons are $1.81 \times 10^{-2}$, $3.14 \times 10^{-3}$, and $3.08 \times 10^{-9}$, respectively, all of which are smaller than $5 \times 10^{-2}$. We therefore reject the null hypothesis that the radii are drawn from the same distribution and conclude that there is a low level of mass segregation present in the cluster within our mass range.

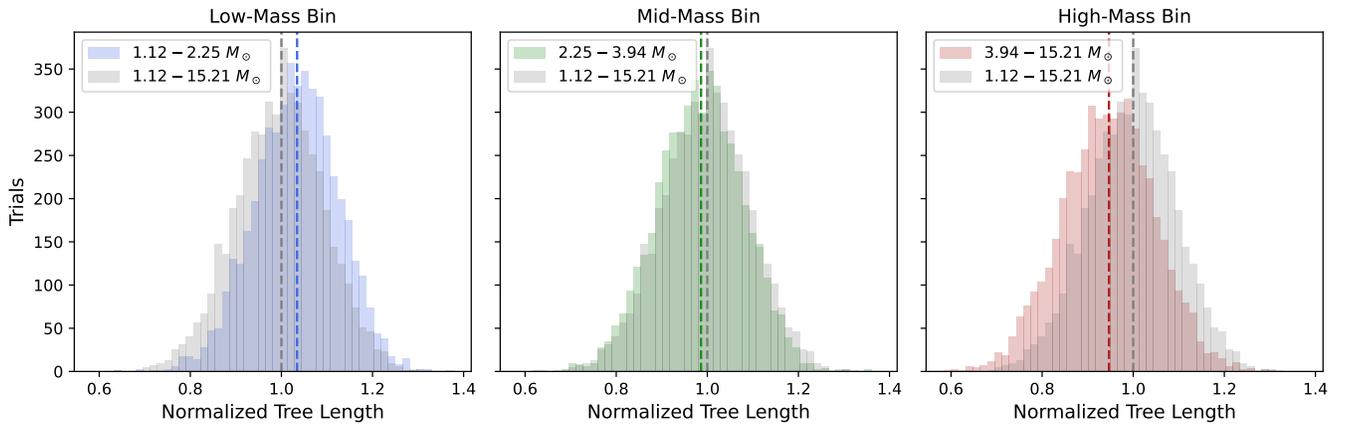

**Figure 20.** Distribution of MST lengths for our low-, mid-, and high-mass bins normalized by the mean MST length of the full catalog. Left: low-mass bin. Middle: mid-mass bin. Right: high-mass bin.

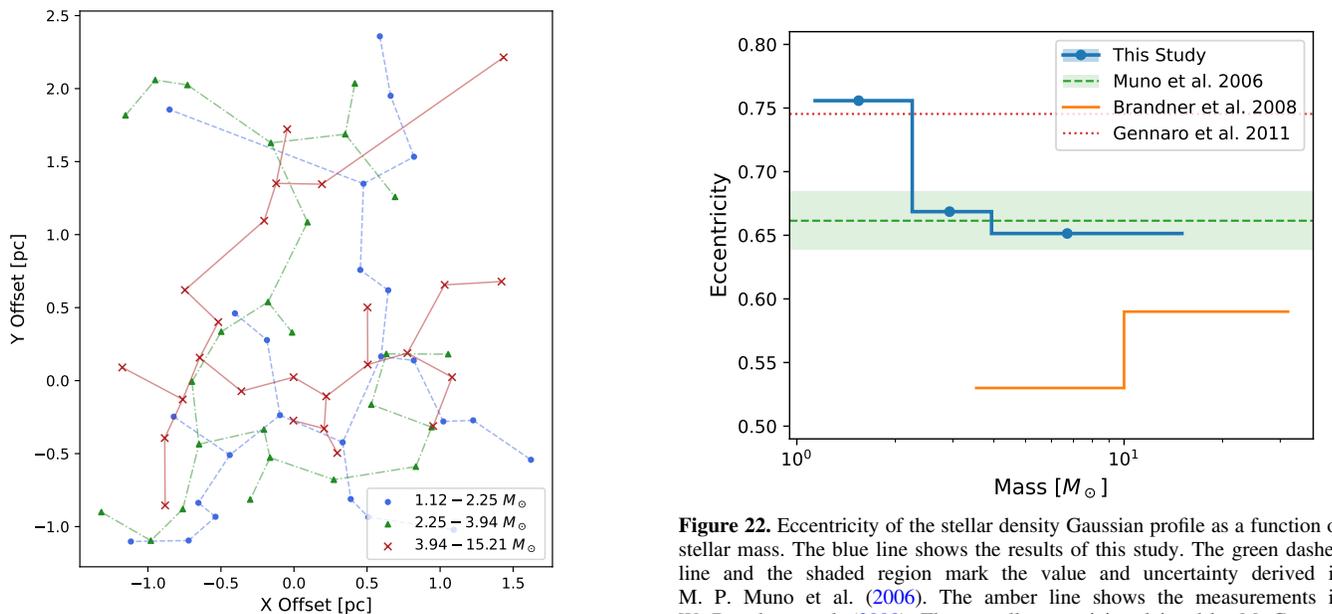

**Figure 21.** MST in each mass bin with the tree length closest to the median normalized tree length in Figure 20. The high-mass stars are marked with red × symbols, the mid-mass stars are marked by green triangles and connected by green dashed lines, and the low-mass objects are marked with blue circles connected by blue dotted lines.

The solar metallicity assumption has little effect on the mass segregation results. With the same supermetallicity test as in Section 4.2 under the MIST model from $[Z] = 0$–$0.3$, the maximum change in the inferred stellar mass is 0.82 $M_\odot$ for a 14.53 $M_\odot$ star, or a 5.6% change. Additionally, the mass increment is mostly proportional to the originally inferred stellar mass. Therefore, we do not anticipate variations in metallicity will alter the relative ordering of stellar masses, which is all that matters in constructing the MSTs. Consequently, the mass segregation results would remain unaffected by metallicity.

## 6. Discussion

With the structural and kinematic information derived in this work, we discuss the virial state, the relevant timescales of the cluster, and their implications.

**Figure 22.** Eccentricity of the stellar density Gaussian profile as a function of stellar mass. The blue line shows the results of this study. The green dashed line and the shaded region mark the value and uncertainty derived in M. P. Muno et al. (2006). The amber line shows the measurements in W. Brandner et al. (2008). The overall eccentricity claimed by M. Gennaro et al. (2011) is marked as the red dotted line.

### 6.1. Elongation

Our measured eccentricity of 0.71 is in good agreement with the value of 0.75 reported by M. Gennaro et al. (2011) and falls within the range of 0.68–0.72 identified by M. Andersen et al. (2017). M. P. Muno et al. (2006) reported a similar value of $e = 0.66 \pm 0.02$ by measuring the diffuse X-ray emissions of the cluster. In addition, we find the eccentricity decreases slightly with increasing mass, with $e = 0.75$, 0.67, and 0.65 for the low-, mid-, and high-mass bins, respectively, a trend that aligns with the findings of M. Gennaro et al. (2011). W. Brandner et al. (2008) measured a moderately smaller $e$ of 0.53 and 0.59 for the mass ranges of 3.5–10 $M_\odot$ and 10–32 $M_\odot$, respectively. The smaller eccentricity may stem from the bias introduced by the authors' assumption about a point-symmetric half-mass radius. Instead, our measurement did not impose any assumption on the mass distribution. Different definitions of eccentricity and ellipticity are employed in the literature. In this work, we adopted the definition of ellipticity and eccentricity as described in Section 4.4. For consistency, we transformed the literature

**Table 5**
Cluster Sample Size by Criteria

| Sample | Constraints | Observed Stars | Weighted Stars[a] |
|---|---|---|---|
| PMs | Detection in All Years | 10,346 | ⋯ |
| PM Box Cut | $v_x, v_y < 3\sigma_{pm}$ | 10,002 | ⋯ |
| Kinematic Probability Cut | $p_\mu \geqslant 0.3$ | 4341 | ⋯ |
| Kinematic and Color Cut | $p_{clust} \geqslant 0.3$ | 3582 | ⋯ |
| Completeness Magnitude Range | $17.5 \geqslant F160W_D \geqslant 11.5$ | 2019 | ⋯ |
| Radial Profile, Full Cluster | $17.50 \geqslant F160W_D \geqslant 11.50$, $r_{eff} \leqslant 2.87$ pc | 1887 | 2304.6 |
| Radial Profile, Low-mass Bin | $17.50 \geqslant F160W_D \geqslant 15.74$ | 762 | 1024.6 |
| Radial Profile, Mid-mass Bin | $15.74 \geqslant F160W_D \geqslant 14.74$ | 627 | 767.6 |
| Radial Profile, High-mass Bin | $14.74 \geqslant F160W_D \geqslant 11.50$ | 498 | 512.4 |
| Radial Profile, Arches Comparison | $4.50\ M_\odot \leqslant M \leqslant 15.21\ M_\odot$ | 396 | 394.2 |
| Radial Profile, Quintuplet Comparison | $2.50\ M_\odot \leqslant M \leqslant 15.21\ M_\odot$ | 1074 | 1210.2 |

**Notes.** The number of stars reflects the cumulative effect of the constraints in the row and all the preceding rows, with the constraints applied sequentially.
[a] The weight refers to $w_i^{r_{eff}}$ in Equation (14) and requires the completeness and area fraction, which is not yet calculated for the first four rows.

values to align with this definition for comparison. Figure 22 shows the comparison of values measured in this study and the literature.

The observed high eccentricity may result from the morphology of the molecular cloud within which Wd1 formed, or imply a hierarchical formation pathway involving the merger of multiple substructures. The decreasing trend of eccentricity with increasing mass can be attributed to the shorter dynamical timescale of more massive stars as they are more concentrated in the center, which can be clearly observed in Figure 13. Therefore, more frequent interactions result in a more isotropic profile. Furthermore, as the elongation aligns with the Galactic plane, this could also arise from the tidal stripping along the plane of less massive stars on the periphery of the cluster.

### 6.2. Virial State

Our measured velocity dispersion of $4.13 \pm 0.13$ km s$^{-1}$ is smaller than the values expected for the cluster if it were in virial equilibrium $\sigma_{vir} = 4.5$–$6.5$ km s$^{-1}$. The discrepancy is about $\sim 3\sigma$, suggesting a subvirial dynamical state. W. Brandner et al. (2008) estimated the virial equilibrium velocity dispersion to be $\sigma_{vir} \geqslant 4.5$ km s$^{-1}$, or 0.25 mas yr$^{-1}$ at their distance estimate of 3.55 kpc. M. Gennaro et al. (2011, 2017) estimate $\sigma_{vir} = 4.5^{+0.8}_{-0.2}$ km s$^{-1}$, and I. Negueruela et al. (2010) reported 6.5 km s$^{-1}$ (see M. Cottaar et al. 2012 for details).

In comparison, S. Mengel & L. E. Tacconi-Garman (2009) measured $\sigma_{1D} = 9.2 \pm 2.5$ km s$^{-1}$ from four RSGs, five yellow hypergiants, and a B-type emission line star. However, this result may be highly overestimated due to the presence of binaries (e.g., M. B. N. Kouwenhoven & R. de Grijs 2008; M. Gieles et al. 2010). M. Cottaar et al. (2012) reported $2.1^{+3.3}_{-2.1}$ km s$^{-1}$, which is derived from the spectroscopic measurements of several PMS stars. Our velocity dispersion measurement is derived from a significantly larger sample with membership and completeness correction. These improvements help mitigate potential biases and enhance the reliability and robustness of our results.

The subvirial state aligns with the claim in M. Cottaar et al. (2012). At the age of Wd1, mass loss due to radiative feedback has already occurred, and the cluster should be dynamically responding. If present, the subvirial state of Wd1 with little gas remaining (e.g., M. G. Guarcello et al. 2025) may imply that the star formation efficiency (SFE; the fraction of the initial gas mass that is turned into stars) is high enough at its formation for it to survive as a bound cluster. Previous studies find that eventually bound clusters are more likely to form in exceptionally high SFE environments, typically greater than 50% (M. P. Geyer & A. Burkert 2001; H. Li et al. 2019). However, there are arguments that up to half of the stars can remain bound with an SFE smaller than 50% (C. M. Boily & P. Kroupa 2003).

Alternatively, cluster formation mechanisms can also explain the gravitationally bound state of Wd1. Instead of forming in a static molecular cloud, stars can form in local overdensities such as gas filaments before they reach the forming star cluster (I. A. Bonnell et al. 2008; S. N. Longmore et al. 2014; L. Wei et al. 2024). Gas expulsion that already happens locally reduces the negative influence of stellar feedback on new star formation, resulting in bound clusters free of gas (J. M. D. Kruijssen 2012; M. G. H. Krause et al. 2020; M. Chevance et al. 2023). The high eccentricity of $e = 0.71$ of Wd1 identified in this work may result from the merging of such local substructures during its formation.

Note that the equilibrium velocity dispersion estimates depend on IMF extrapolations, which need further confirmation. The viral velocity estimate would benefit from a more complete sample of low-mass objects in Wd1. Distance is another factor affecting the virial state. Assuming the current estimate of virial equilibrium velocity dispersion, Wd1 would need to be at a distance of 4.0–5.8 kpc instead of 3.7 kpc for the observed velocity dispersion to match the virial equilibrium model. The lower bound of distance is compatible with the literature estimates and the spread in the posterior caused by degeneracy as reported in Section 5.3. Though the subvirial state of Wd1 cannot be definitively confirmed in the presence of the uncertainties in the IMF extrapolation and distance estimates, the distance modeled in this work and the literature values of 4.0–4.5 kpc all put the cluster closer to the subvirialized or virialized state.

### 6.3. Radial Profile Comparisons

We compare the radial density profile of Wd1 with those of the Arches and Quintuplet clusters, modeled using the same methodology. We set a minimum mass limit of 4.5 $M_\odot$ and 2.5 $M_\odot$ to keep consistent mass ranges with Arches and Quintuplet, respectively. Details regarding the numbers of unweighted and weighted stars per bin and bin edges in mass are provided in Table 5.

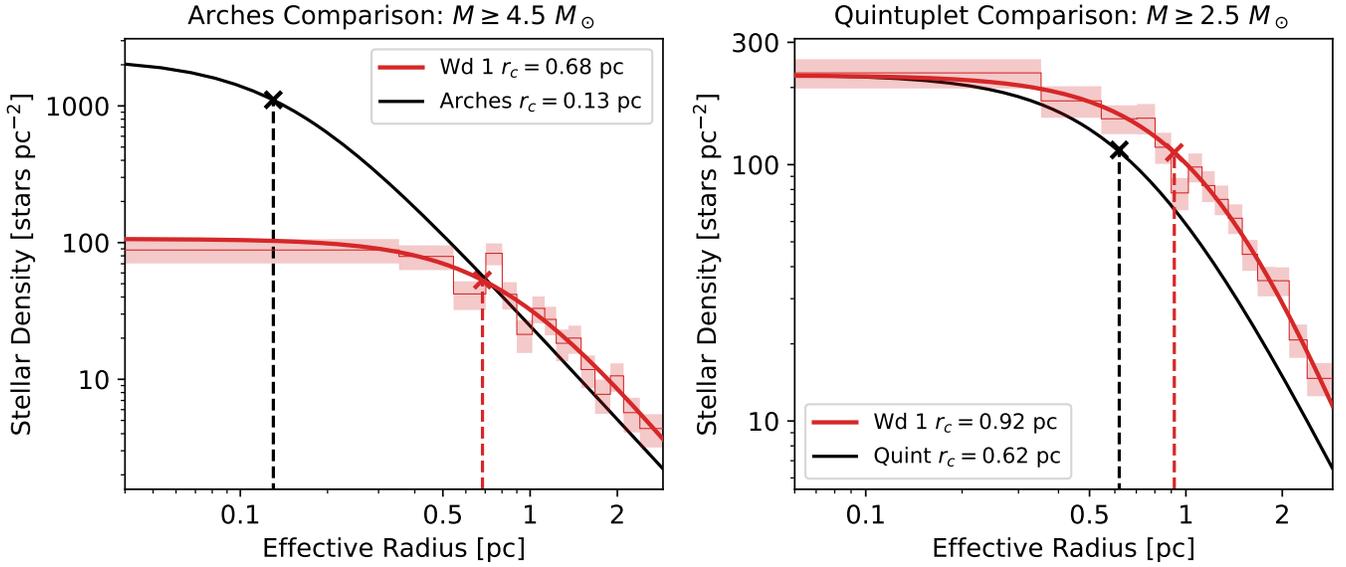

**Figure 23.** Comparison of the radial profile of Wd1 to the Arches and Quintuplet clusters using the EFF fit parameters in M. W. Hosek et al. (2015) and N. Z. Rui et al. (2019). We restrict the profile fit to $M \geqslant 4.5\ M_\odot$ and $M \geqslant 2.5\ M_\odot$ for the Arches and Quintuplet comparison, respectively.

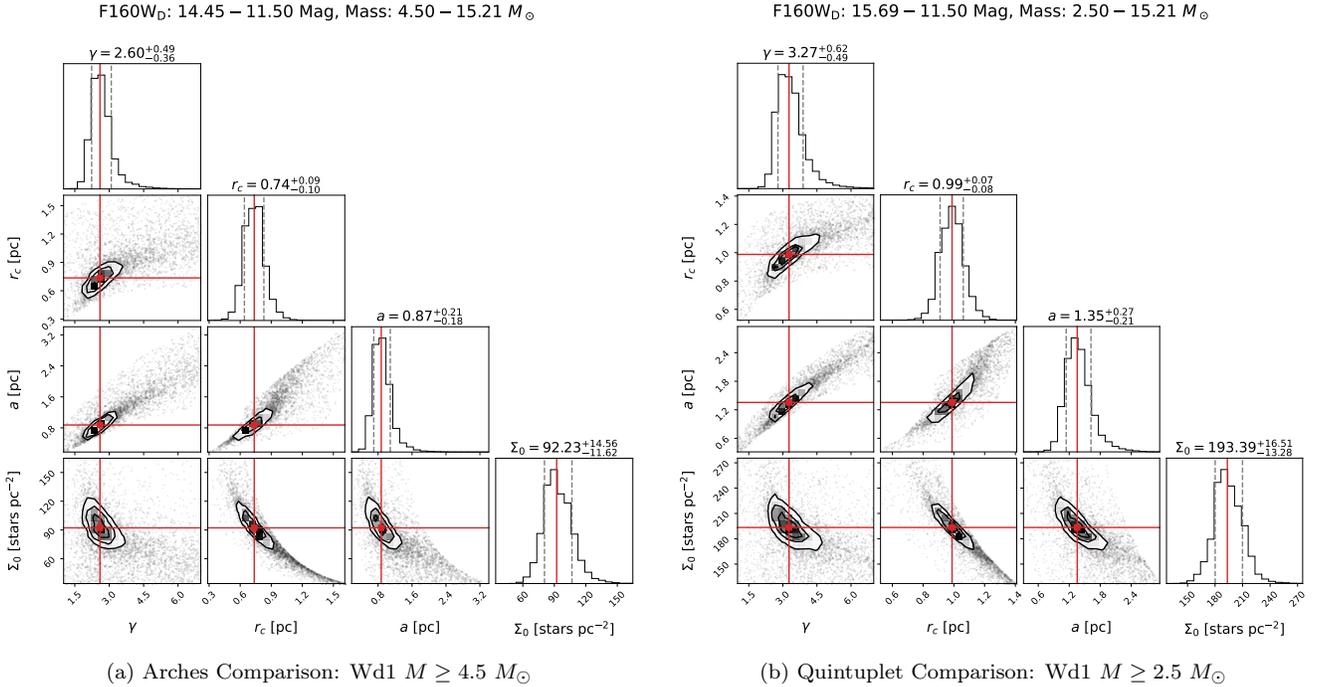

(a) Arches Comparison: Wd1 $M \geq 4.5\ M_\odot$

(b) Quintuplet Comparison: Wd1 $M \geq 2.5\ M_\odot$

**Figure 24.** Weighted posterior distributions of EFF radial profile parameters. (a) Sources with mass greater than 4.5 $M_\odot$ for comparison with the Arches cluster. (b) Sources with mass greater than 2.5 $M_\odot$ for comparison with the Quintuplet cluster. The red lines mark the weighted median, and the gray dashed line marks the weighted 16th and 84th percentiles. Note that we only modeled $r_c$, and the posterior distribution of $a$ is purely converted from $r_c$.

Figure 23 shows the EFF radial profile comparisons with Arches and Quintuplet, respectively. For consistency of comparisons, the masses are restricted to $\geqslant 4.5\ M_\odot$ and $\geqslant 2.5\ M_\odot$. The associated posteriors are illustrated in Figure 24. The Arches cluster displays a stellar density more than 10 times denser than Wd1 in the core, while the Quintuplet cluster shares a similar density profile and core radius with Wd1.

### 6.4. Dynamical Timescales

In this section, we discuss the crossing time, relaxation time, the expected mass segregation timescales, and their implications on whether the mass segregation in Wd1 is dynamical or primordial.

The crossing time of Wd1 is

$$t_{\rm cross} = \frac{r_{\rm hm}}{\sigma_{\rm 1D}} = 0.26\ {\rm Myr}, \qquad (33)$$

where we adopted the half-mass radius of $r_{\rm hm} = 1.09$ pc corrected by completeness and area fraction, after excluding stars with an area fraction $f < 0.3$ used in modeling the velocity dispersion as in Section 5.4. This is consistent with the half-mass radius used in W. Brandner et al. (2008). In

**Table 6**
Elson–Fall–Freeman Radial Profile Results

| Sample | $M$ ($M_\odot$) | F160W$_D$ (mag) | $r_{\rm eff}$[a] (pc) | $\gamma$[b] | $r_c$[c] (pc) | $a$ (pc) | $\Sigma_0$ (stars pc$^{-2}$) |
|---|---|---|---|---|---|---|---|
| Full Cluster | 1.12–15.21 | 17.50–11.50 | ⩽2.87 | $2.85^{+0.41}_{-0.35}$ | $1.02 \pm 0.06$ | $1.29^{+0.19}_{-0.17}$ | $352.05^{+20.94}_{-19.59}$ |
| Low-mass Bin | 1.12–2.25 | 17.50–15.74 | ⩽2.87 | $3.31^{+1.30}_{-0.77}$ | $1.32^{+0.10}_{-0.11}$ | $1.84^{+0.54}_{-0.41}$ | $114.93^{+9.13}_{-7.71}$ |
| Mid-mass Bin | 2.25–3.94 | 15.74–14.74 | ⩽2.87 | $3.63^{+1.18}_{-0.77}$ | $1.03 \pm 0.10$ | $1.52^{+0.42}_{-0.33}$ | $122.34^{+13.06}_{-10.64}$ |
| High-mass Bin | 3.94–15.21 | 14.74–11.50 | ⩽2.87 | $2.67^{+0.50}_{-0.38}$ | $0.74 \pm 0.08$ | $0.89^{+0.20}_{-0.17}$ | $123.78^{+16.94}_{-13.50}$ |
| Arches Comparison | 4.5–15.21 | 14.45–11.50 | ⩽2.87 | $2.60^{+0.49}_{-0.36}$ | $0.74^{+0.09}_{-0.10}$ | $0.87^{+0.21}_{-0.18}$ | $92.23^{+14.56}_{-11.62}$ |
| Quint Comparison | 2.5–15.21 | 15.69–11.50 | ⩽2.87 | $3.27^{+0.62}_{-0.49}$ | $0.99^{+0.07}_{-0.08}$ | $1.35^{+0.27}_{-0.21}$ | $193.39^{+16.51}_{-13.28}$ |

**Notes.** Description of columns. Sample: star sample used in the corresponding analysis. M: mass range of each sample. F160W$_D$: dereddened magnitude in F160W. $r_{\rm eff}$: effective radius as defined in Equation (12). $\gamma$: slope of the radial profile power law in Equation (15). $r_c$: core radius in Equation (16). $a$: core parameter converted from $r_c$ and not sampled. $\Sigma_0$: amplitude of the radial profile in Equation (15).
[a] Radius limit set by the distance at which the area fraction $f \geqslant 0.3$.
[b] Uniform prior $U(1, 6)$.
[c] Uniform prior $U(0, 2)$.

**Table 7**
Part of the Color–Magnitude Diagram Fitting Parameters and Results

| Parameter | Description | Prior | Result | Unit |
|---|---|---|---|---|
| $\log_{10} t$ | Age | $U(6.3, 7.3)$ | $6.87 \pm 0.03$ | $\log_{10}$(yr) |
| $d$ | Distance | $U(3, 5)$ | $3723.8 \pm 113.3$ | kpc |
| $A_{Ks}$ | Average extinction | $U(0.7, 0.76)$ | $0.72 \pm 0.00$ | mag |
| $dA_{Ks}$ | Differential extinction | $U(0, 0.09)$ | $0.05 \pm 0.00$ | mag |

comparison, M. Gennaro et al. (2017) estimated a crossing time of 0.2 Myr assuming Wd1 is virialized with a full radius of 2 pc. Our estimate results directly from the velocity dispersion measurements, without any assumption about the virial state. The age of Wd1 is therefore about 36 times its crossing time.

The relaxation time can be calculated from the number of stars $N$ and the crossing time $t_{\rm cross}$ (J. Binney & S. Tremaine 2008):

$$t_{\rm relax} = \frac{N}{10 \ln N} t_{\rm cross} = 0.22 \text{ Gyr}, \quad (34)$$

where we assume the total number of stars is $N \sim 10^5$ (W. Brandner et al. 2008; M. Gennaro et al. 2017). This is about 30 times the age of Wd1.

The timescale for a star of mass $M$ to reach energy equipartition and therefore dynamical mass segregation is

$$t(M) \sim \frac{\overline{m}}{M} t_{\rm relax}, \quad (35)$$

where $\overline{m}$ is the average mass of the cluster. Considering the cluster age of 7.45 Myr and an average mass of $\sim 0.4\, M_\odot$ (M. Gennaro et al. 2017), we would expect mass segregation for stars more massive than $\sim 12\, M_\odot$. In comparison, M. Cottaar et al. (2012) estimated a segregation mass of $20\, M_\odot$, and M. Gennaro et al. (2017) reported little or no mass segregation except for the $>40\, M_\odot$ stars in Wd1 based on observations. Recall that our analysis is restricted to stars in the mass range of 1.12–15.21 $M_\odot$, which accounts for the low-level segregation observed in this study. With a mass segregation ratio of $\Lambda_{\rm MSR} = 1.11 \pm 0.11$, it only marginally greater than one by 1$\sigma$. This is consistent with the expectation from the timescale analysis that the stellar population in this mass range does not exhibit strong segregation.

The consistency between the observed minor mass segregation in the lower-mass range, combined with the segregation mass estimates based on dynamical timescales, implies that the segregation in Wd1 is likely dynamical rather than primordial, in agreement with previous studies (M. Gennaro et al. 2017). Considering its age and relaxation timescale, the dynamical process in this mass range is possibly still ongoing. The highest-mass stars are already segregated, and lower-mass stars are beginning to experience the effects of dynamical processes, therefore displaying a weak sign of mass segregation.

## 7. Conclusion

We analyze the kinematics and structure of Wd1 using multiepoch astrometric and photometric data from HST WFC3-IR filters. We model the kinematic and color memberships of the stars. The structure of Wd1 is thoroughly analyzed after correcting for membership, extinction, and completeness. Specifically, we report the following conclusions.

i. We obtained the cluster membership of 10,346 observed stars, consisting of PM kinematic membership characterized by a three-component GMM model, a Boolean photometric membership, and completeness correction.

ii. We construct a stellar density map of Wd1 corrected by a spatial reddening map and cluster membership of each star.

iii. With the stellar density map, we find that the cluster is elongated in the northeast–southwest direction, with an eccentricity of 0.71 and the semimajor axis is at a position angle of $\sim 35.°2$ east of north. The elongation aligns with the Galactic plane. Furthermore, eccentricity decreases with increasing mass.

iv. The high eccentricity may be inherited from the molecular cloud from which Wd1 formed, or imply a formation process during which multiple substructures merged. The alignment of the elongation with the Galactic plane, coupled with the higher eccentricity and more spatially diffuse distribution of low-mass stars, may also indicate tidal disruption within the Galactic plane.

v. We fit an EFF radial profile model to the stellar density and observed a slight decrease in the core radius with increasing stellar mass, indicative of minor mass segregation.

vi. Another weak sign of mass segregation is identified by comparing the MST length for different mass ranges, with a relatively small mass segregation ratio of $\Lambda_{\rm MSR} = 1.11 \pm 0.11$.

vii. We present the velocity dispersion measurements for 1211 stars. The velocity dispersion is $(\sigma_{R,\rm cl}, \sigma_{T,\rm cl}) = (0.25 \pm 0.01, 0.22 \pm 0.01)$ mas yr$^{-1}$, translating into $(4.33 \pm 0.19, 3.91 \pm 0.16)$ km s$^{-1}$. The 1D velocity dispersion of $\sigma_{\rm 1D} = 4.13 \pm 0.13$ km s$^{-1}$ is slightly below the virial equilibrium estimate of $\sigma_{\rm vir} = 4.5$–$6.5$ km s$^{-1}$ reported in the literature, suggesting the cluster is on the verge of subvirializing. This conclusion is subject to uncertainties in the distance and IMF extrapolations.

viii. The subvirial, gravitationally bound state of Wd1 with little gas remaining implies either an exceptionally high SFE, likely >50% at its formation, or it formed from the merging of substructures like gas filaments that already started local gas expulsion driven by stellar feedback before they reach the cluster.

ix. The crossing time is 0.26 Myr with a mean projected cluster radius of $r_{\rm hm} = 1.09$ pc weighted by membership, completeness, and area fraction. The age of Wd1 of 7.45 Myr is about 29 times its crossing time. The relaxation time is 0.22 Gyr, about 30 times its age.

x. Given the age and relaxation time, we expect mass segregation for stars down to 12 $M_\odot$, which accounts for the weak sign of segregation in our analysis with $\Lambda_{\rm MSR} = 1.11 \pm 0.11$, as this work is restricted to sources between the mass range of 1.12–15.21 $M_\odot$. This implies that mass segregation is more likely dynamical rather than primordial in Wd1.

## Acknowledgments

The HST data presented in this article were obtained from the Mikulski Archive for Space Telescopes (MAST) at the Space Telescope Science Institute. The specific observations analyzed can be accessed via DOI:10.17909/r5kb-7h31. We acknowledge the anonymous referee for providing constructive feedback. L.W. acknowledges Prof. Smadar Naoz for providing valuable advice on timescale analysis. This research is based on observations made with the NASA/ESA Hubble Space Telescope obtained from the Space Telescope Science Institute (STScI), which is operated by the Association of Universities for Research in Astronomy, Inc., under NASA contract NAS 5–26555. These observations are associated with programs GO-13044 and GO-13809. J.R.L., P.C.B., D.K., and M.S. acknowledge support from the National Science Foundation Astronomy and Astrophysics Grant AST-1764218. N.Z.R. acknowledges support from the National Science Foundation Graduate Research Fellowship under grant No. DGE–1745301. Portions of this work were conducted at the University of California, San Diego, which was built on the unceded territory of the Kumeyaay Nation, whose people continue to maintain their political sovereignty and cultural traditions as vital members of the San Diego community.

*Facility:* HST WFC3.

*Software:* astropy (Astropy Collaboration et al. 2013, 2018, 2022), emcee (D. Foreman-Mackey et al. 2013), Matplotlib (J. D. Hunter 2007).

## ORCID iDs

Lingfeng Wei (魏凌枫) ● https://orcid.org/0000-0002-2612-2933
Peter C. Boyle ● https://orcid.org/0000-0001-8513-2608
Jessica R. Lu ● https://orcid.org/0000-0001-9611-0009
Matthew W. Hosek, Jr. ● https://orcid.org/0000-0003-2874-1196
Quinn M. Konopacky ● https://orcid.org/0000-0002-9936-6285
Richard G. Spencer ● https://orcid.org/0000-0001-7101-4328
Dongwon Kim ● https://orcid.org/0000-0002-6658-5908
Nicholas Z. Rui ● https://orcid.org/0000-0002-1884-3992
Jay Anderson ● https://orcid.org/0000-0003-2861-3995